\documentclass[%
 preprint,
 amsmath,amssymb,
 aps,
]{revtex4-2}

\usepackage{graphicx}% Include figure files
\usepackage{dcolumn}% Align table columns on decimal point
\usepackage{bm}% bold math
\usepackage[utf8]{inputenc} % allow utf-8 input
\usepackage[T1]{fontenc}    % use 8-bit T1 fonts
\usepackage{hyperref}       % hyperlinks
\usepackage{url}            % simple URL typesetting
\usepackage{booktabs}       % professional-quality tables
\usepackage{amsfonts}       % blackboard math symbols
\usepackage{nicefrac}       % compact symbols for 1/2, etc.
\usepackage{microtype}      % microtypography
\usepackage{lipsum}		% Can be removed after putting your text content
\usepackage{graphicx}

\usepackage[english]{babel}
\usepackage{amsmath}
\usepackage{relsize}
\usepackage{chngcntr}
\usepackage{nameref,hyperref}
\usepackage{amssymb}
\usepackage{subcaption}
\usepackage[verbose]{placeins}
\usepackage{graphicx}
\usepackage{color}
\usepackage{multirow}
\usepackage{wasysym}
\usepackage[symbol]{footmisc}
\usepackage{tabulary}
\usepackage{chngcntr}
\counterwithout{figure}{section}
\usepackage{amsmath,amsfonts,amsthm,bm} % Math packages

\usepackage{changes}
\begin{document}

%\preprint{APS/123-QED}

\title{Sampling rate-corrected analysis of irregularly sampled time series}

\author{Tobias Braun}
 \email{tobraun@pik-potsdam.de}
 \affiliation{Potsdam Institute for Climate Impact Research (PIK)\\
	Member of the Leibniz Association\\
	14473 Potsdam, Germany\\
	Tel.: +49-331-28820744 \\}

\author{Cinthya N. Fernandez}
\affiliation{Institute for Geology, Mineralogy and Geophysics\\
	Ruhr-Universität Bochum\\
	44801 Bochum\\
	Germany}

\author{Deniz Eroglu}
\affiliation{Faculty of Engineering and Natural Sciences\\
	Kadir Has University\\
	34083 Istanbul\\
	Turkey}

\author{Adam Hartland}
\affiliation{Environmental Research Institute, School of Science\\ University of Waikato\\
Hamilton, Waikato 3240\\
New Zealand}

\author{Sebastian F. M.~Breitenbach}
\affiliation{	Department of Geography and Environmental Sciences\\
	Northumbria University\\
	Newcastle upon Tyne, NE1 8ST, UK}

\author{Norbert Marwan}
\affiliation{Potsdam Institute for Climate Impact Research (PIK)\\
	Member of the Leibniz Association\\
	14473 Potsdam, Germany}
\altaffiliation[Also at: ]{University of Potsdam\\
	Institute of Geosciences\\
	14473 Potsdam, Germany}

\date{\today}

\begin{abstract}
The analysis of irregularly sampled time series remains a challenging task requiring methods that account for continuous and abrupt changes of sampling resolution without introducing additional biases. The edit--distance is an effective metric to quantitatively compare time series segments of unequal length by computing the cost of transforming one segment into the other. 
We show that transformation costs generally exhibit a non-trivial relationship with local sampling rate. If the sampling resolution undergoes strong variations, this effect impedes unbiased comparison between different time episodes.
We study the impact of this effect on recurrence quantification analysis, a framework that is well-suited for identifying regime shifts in nonlinear time series. A constrained randomization approach is put forward to correct for the biased recurrence quantification measures. This strategy involves the generation of a novel type of time series and time axis surrogates which we call sampling rate constrained (SRC) surrogates. We demonstrate the effectiveness of the proposed approach with a synthetic example and an irregularly sampled speleothem proxy record from Niue island in the central tropical Pacific. Application of the proposed correction scheme identifies a spurious transition that is solely imposed by an abrupt shift in sampling rate and uncovers periods of reduced seasonal rainfall predictability associated with enhanced ENSO and tropical cyclone activity.
\end{abstract}


\maketitle

\section{Introduction}
\label{sec1}
The analysis of time series from complex systems calls for numerical methods that capture the 
most relevant features in the observed variability. At the same time, the impact of various 
frequently encountered data-related intricacies such as low signal-to-noise ratio, 
nonstationarity, and limited time series length must be accounted for. A major challenge is 
posed by irregular sampling, i.e., variations in the interval $\Delta_i = t_i - t_{i-1}$ 
between consecutive measurement times
$t_{i-1}$ and $t_i$. Irregular sampling is observed in many complex real-world systems. The underlying mechanisms that render the temporal sampling irregular may differ: sampling can be inherently 
irregular due to an additional process that controls the sampling interval (e.g., financial or 
cardiac time series \cite{reno2003closer, fedotov2020methods}); a mixture of various external 
processes can result in `missing values', i.e., multiple interacting processes result in the 
non-availability of measurements (e.g., sociological or psychological survey data 
\cite{enders2010applied}) or cause failures of the system (e.g., mechanical/electronical 
systems \cite{khayati2020mind}); finally, the measurement process often results in irregularly 
sampled time series (e.g., astronomical \cite{scargle1982studies} or geophysical systems 
\cite{rehfeld2011comparison}). Proxy time series obtained from 
palaeoclimate archives are a particularly challenging example since irregularity in the 
temporal sampling can itself contain valuable information on the processes of interest \cite{baldini2021detecting}. The 
growth rate of a stalagmite for example depends on variable environmental factors, including temperature in 
the cave and drip rate \cite{muhlinghaus2007modelling}, among others. Since these factors and their variability are strongly coupled to the environmental conditions outside the cave, growth rate must be regarded as a dynamical indicator for example, hydrological conditions which in turn determine variations in 
the temporal sampling of the proxy time series.

Across many research communities, resampling based on interpolation techniques and imputation 
approaches are popular methods for making irregularly sampled time series compatible with standard time 
series analysis tools \cite{garland2018anomaly, trauth2021spectral}. Artefacts and statistical biases caused by interpolation techniques are well-known and may result in misinterpretation of the extracted time series properties, an issue further aggravated by the fact that biases introduced by interpolation may vary among different 
systems \cite{rehfeld2013late}. The robustness of results arising from different 
interpolation techniques for the same data set is rarely examined. For instance, linear 
interpolation will not compensate for the effect of lower variability during sparesely sampled 
episodes in a time series compared to more densely sampled periods. In fact, linear 
interpolation and mean imputation decrease variance to a hardly quantifiable, data-related degree \cite{schulz1997spectrum}.
Finally, more complex imputation models may account for such finite (sampling) size effects 
but may not represent the `natural' variability of a time series adequately. For data 
\textit{not-missing-at-random}, the assignment of a sufficient imputation model can be challenging and 
must account for nonstationarity in the underlying non-random effects (e.g., for the palaeoclimate example mentioned above). Similar biases are known from the problem of imbalanced data, i.e., given two populations that should be compared based on a statistical model, a majority class exists that contains significantly more samples than the minority class and thus, oversampling techniques are applied to compensate for the resulting bias \cite{chawla2002smote, barua2012mwmote}.

Geophysical time series frequently exhibit nonlinear features such as nonlinear oscillations 
and critical regime transitions, e.g., tipping points \cite{lenton2019climate}. Dynamical 
system theory regards observations from such systems as embedded in a higher-dimensional phase 
space and offers a range of tools to quantify gradual or abrupt changes in these 
dynamics  \cite{bradley2015nonlinear, marwan2021nonlinear}. The power of these methods relies on their ability to 
uncover features that regular techniques, such as autocorrelations or 
variance estimation, fail to uncover \cite{scheffer2009early}.
Aiming for higher applicability of nonlinear time series analysis methods in the Earth 
sciences, irregular sampling approaches have been proposed 
\cite{lekscha2018phase, mccullough2016counting, sakellariou2016counting}. One of these approaches is based on the idea of transforming sub-sequences of unequal lengths in a time series into each other and comparing the costs of these transformations for all sub-sequences 
\cite{suzuki2010definition}.
More generally, the definition of a metric distance between states at different instances of 
time can entail dynamical information on the evolution of the phase space trajectory of the 
studied system.
While standard metrics (such as Euclidean distance) fail to account for irregular sampling, 
the TrAnsformation-Cost Time-Series (TACTS)  \cite{ozken2015transformation} includes the 
temporal information for distinct time series segments. Similar approaches based on the edit 
or Levenshtein distance have been used in natural language processing 
\cite{ukkonen1985algorithms} and metric analyses of point processes 
\cite{banerjee2020recurrence}, among many others.

In this work, we focus on the application of the (m)Edit-distance \cite{ozken2018recurrence} as a base for the 
computation of recurrence plots (RPs) \cite{marwan2007recurrence}. The (m)Edit-distance 
approach can potentially be employed in any methodological framework that includes computation 
of a distance (or similarity) measure. The RP technique represents one particular application that has proven to be a powerful 
approach, tackling many of the fundamental problems in time series analysis, such as time 
series classification \cite{garcia2018classification}, the study of synchronization between 
multiple time series \cite{romano2004multivariate}, and detection of regime transitions 
\cite{marwan2013recurrence}. Recurrence quantification analysis (RQA) provides a means of 
quantifying the tendency of a time series to revisit previously visited states and has grown 
in its scope from basic predictability quantification towards more ambitious measures that, 
e.g., capture the multiscale nature of transitions \cite{corso2018quantifying, 
schinkel2007order, braun2021detection}.
The identification of shifts stands out as a 
particularly interesting application since critical transitions can often be linked to the 
vulnerability of the respective regional climate system towards external shocks or feedback 
mechanisms. The combination of the (m)Edit-distance approach and RPs offers a promising 
approach to identify regime transitions in irregularly sampled records, which may otherwise be impeded 
without an adequate technique designed to account for sampling variations 
\cite{ozken2018recurrence, marwan2018regime, stegner2019inferring}. In following this approach, 
special care must be taken if irregular sampling intervals undergo strong variations, i.e., where the process(es) that control the sampling rate are rendered non-stationary. In some 
applications, segments can be chosen such that they do not cover the same time period but the same number of values on average. Other applications 
require fixing a particular time period to be covered by each segment since this time period corresponds to the time scale under investigation, e.g., a year for seasonal time series. 
Even if such an approach is not motivated by the research question, splitting the time series 
into segments that correspond to non-equal time periods will result in mixing of time scales 
in the resulting distance matrix if the sampling rate is highly non-stationary. 
Here, we focus on segments that cover equal time periods but varying numbers of values, refered 
to as \textit{segment size}. We will show that in such cases, the resulting strong variations in segment size entail a 
non-trivial sampling bias of the (m)Edit-distance.

We introduce the (m)Edit-distance methodology in 
Sect.~\ref{sec2.1} followed by a short summary of recurrence analysis in Sect.~\ref{sec2.2}. 
Sect.~\ref{sec3} illustrates the problem of strong variations in the sampling rate whereas model time series are studied to elucidate the sample size effects. A correction 
scheme based on constrained randomization is proposed in Sect.~\ref{sec4}. 
In Sect.~\ref{sec5}, we demonstrate the importance to correct for the identified sample-size dependence in an application to a palaeoclimate record from Niue island in the central Pacific where we identify variations in seasonal predictability. We conclude our findings 
in Sect.\ref{sec6}.

\section{Methodology}
\label{sec2}
\subsection{The (m)Edit-distance measure}
\label{sec2.1}
Many approaches in nonlinear time series analysis are based on some notion of a 
(dis)similarity measure. For deterministic systems, embedding the univariate time series into 
an $m$-dimensional phase space offers a multitude of quantitative approaches to analyse the 
variability of its trajectory \cite{kraemer2021unified}. Yet appropriate techniques to extract the embedding dimension and delay from empirical data are needed. These approaches can be cumbersome. In this work we focus 
on univariate time series wherein the most widespread dissimilarity measure 
between distinct segments $\mathcal{S}_{a},\, \mathcal{S}_{b}$ is the Euclidean distance. It 
is a metric distance, i.e., its value is always positive $D(\mathcal{S}_{a},\, 
\mathcal{S}_{b}) 
\, \geq \, 0$, it is symmetric $D(\mathcal{S}_{a},\, \mathcal{S}_{b}) \, = \, 
D(\mathcal{S}_{b},\, \mathcal{S}_{a})$, and the triangle inequality holds 
$D(\mathcal{S}_{a},\, 
\mathcal{S}_{\mathrm{c}}) \, \leq D(\mathcal{S}_{a},\, \mathcal{S}_{b}) + 
D(\mathcal{S}_{b},\, 
\mathcal{S}_{\mathrm{c}})$. If the time series is characterized by missing values or the 
sampling interval $\Delta_i$ is irregular (e.g., due to irregularities in the measurement process), no straight-forward application of Euclidean distance or comparable metrics is possible: dissimilarity of values at unequal time scales would be computed without accounting for their non-equality.
Linear interpolation as a means of resampling the time series values onto a regular time axis is among the most popular 
approaches to regularize sampling \cite{rehfeld2014similarity}. Yet, hardly controllable artefacts arise from linear interpolation, ranging from difficulties related to altered absolute timing to underestimation of variance or overestimation of persistence \cite{rehfeld2013late, 
mudelsee2013climate}.

Originally proposed for natural language processing, the edit distance measure 
\cite{masek1980faster} is designed to compare sequences of variable length. 
Shifting and adding $\&$ deleting of strings were proposed as two elementary operations to 
quantify dissimilarities between words, an objective also pursued by other methods such as 
dynamic time warping \cite{rabiner1978considerations}. The resulting costs are calculated by 
identifying a minimum cost path to transform one sequence into the other. Taking the next step 
towards an application to empirical time series, the edit distance was applied to point 
process data whereby cost parameters for the elementary operations remained arbitrary 
\cite{suzuki2010definition, victor1997metric}. By equipping the technique with data-driven 
cost parameter estimates, it was then applied to irregularly sampled palaeoclimate time series 
\cite{ozken2015transformation}. A further modification ((m)Edit distance) with an application 
to extreme events was proposed to consider the saturation of shifting costs when a 
certain time scale $\tau$, separating the two compared segments, is exceeded 
\cite{banerjee2020recurrence}. The main difference between applying the edit distance to 
series of events/spike trains and irregularly sampled time series is that for the latter, 
amplitudes of time series values must be considered. In the following, whenever no assumptions are 
made about the amplitudes of a signal, we refer to `events'.
The edit distance between two segments $\mathcal{S}_{a},\, \mathcal{S}_{b}$ of an irregularly 
sampled time series is computed by minimizing the transformation costs by:
\begin{widetext}
\begin{align}
D(\mathcal{S}_{a},\, \mathcal{S}_{b}) \, = \, {\mathrm{min}}
\left\{
\sum_{\alpha,\beta \in \mathcal{C}} 
\left[
\underbrace{f_{\Lambda_0}\left(t(\alpha), t(\beta);\,\tau\right)}_\text{shifting}
\, + \, 
\underbrace{\Lambda_{\mathrm{k}}\bigl\|L_{a}(\alpha) - L_{b}(\beta)\bigr\|}_\text{amplitude change}
\right]
\ + \ 
\underbrace{\Lambda_{\mathrm{S}}\left(|I| + |J| - 2|\mathcal{C}|\right)}_\text{adding and deleting}
\right\}
\label{meq1}
\end{align}
\end{widetext}
with a norm $\|\cdot\|$ (e.g., the Euclidean 
norm), the $\alpha\,$-th/$\beta\,$-th amplitudes $L_{a}(\alpha),\, L_{b}(\beta)$ of the 
segments $\mathcal{S}_{a},\, \mathcal{S}_{b}$ and the cardinalities $|\cdot|$ of the sets $I,\, J 
\,$ and $\mathcal{C}$. While the latter are a  set  of  indices  of  the time series values, $\mathcal{C}$ denotes the values that are shifted.
$D(\mathcal{S}_{a},\, \mathcal{S}_{b})$ is a metric distance. The cost 
parameters $\Lambda_0,\,\Lambda_{\mathrm{k}}\, $, and $\Lambda_{\mathrm{S}}$ need to be fixed 
prior to cost optimization.
We choose the cost parameter for amplitudes changes $\Lambda_{\mathrm{k}}$ as suggested in \cite{ozken2015transformation}:
\begin{align}
\Lambda_{\mathrm{k}} \, = \, \frac{M-1}{\sum_{i=1}^{M-1}\|x_i - x_{i+1}\|}
\label{meq2}
\end{align}
The cost parameter $\Lambda_{\mathrm{S}}$ for deleting $\&$ adding has to be chosen such that 
deletions are neither `too cheap' nor `too expensive'. For a set of time series values with a 
large temporal distance or very distinct amplitudes, a deletion and addition should be favorable while a 
too low value of $\Lambda_{\mathrm{S}}$ will result in a transformation of sequences solely by 
deletion and adding operations even for very close time series values. We follow the scheme proposed in \cite{ozken2018recurrence} by 
assuming normality for the distance values between all segments of the time series and 
optimize $\Lambda_{\mathrm{S}}$ within a specified range using a Kolmogorov-Smirnov (KS)-test to ensure that the normality assumption holds as close as possible.
Following the modification proposed in \cite{banerjee2020recurrence}, costs associated with 
shifting of time instances between two time series values are controlled by the logistic 
function
\begin{align}
f_{\Lambda_0}\left(t(\alpha), t(\beta);\,\tau\right) \, = \, \frac{\Lambda_0}{1 + \mathrm{e}^{-\left(\|t_{a}(\alpha) - t_{b}(\beta)\| - \tau\right)}}
\label{meq3}
\end{align}
where $\tau$ is the location parameter of the logistic function, reflecting a 
characteristic time scale that separates exponentially increasing from saturating/bounded 
exponentially increasing costs for shifting. We choose $\tau$ as 
the average sampling interval of the time series; $\tau \, = \, \left. T \middle/ M \right.$  
with the total time period $T$ and the number of samples $M$. Interpreting $\tau$ as a `temporal 
tolerance', this choice ensures that shifting exponentially fast becomes less favorable if 
time instances are separated by several standard deviations of the sampling interval 
distribution. Finally, a value for the maximum costs associated with shifting $\Lambda_0$ needs to be 
set. The ratio $\left. \Lambda_{\mathrm{K}}\middle/\Lambda_0 \right.$ reflects the relative 
importance of temporal and magnitudinal separation; in the limiting case $\left. 
\Lambda_{\mathrm{K}}\middle/\Lambda_0 \right. \,\gg\, 1\,$, irregular sampling is no longer 
accounted for and the resulting distance between two segments solely 
reflects the norm $||L_{a}(\alpha) - L_{b}(\beta)||$
for all amplitudes $L_{a}(\alpha),\, L_{b}(\beta)$ of both segments $\mathcal{S}_{a},\, 
\mathcal{S}_{b}$.
In the opposite case $\left. \Lambda_{\mathrm{K}}\middle/\Lambda_0 \right. \,\ll\, 1\,$, the 
time series can be regarded as a series of events since cost optimization is independent of 
their amplitudes. We choose $\Lambda_{\mathrm{K}} = \Lambda_0 = 1$. It must be stressed that this rate depends on the research question and the data under study.

In the following, we discuss the finite-sample effects bias (m)Edit-distance values 
$D(\mathcal{S}_{a},\, \mathcal{S}_{b})$ and give a summary of the RP methodology. This facilitates the presentation of finite-sample effects discussed in 
Sect.~\ref{sec3} alongside an illustration of the (m)Edit-distance 
methodology (Fig.~\ref{fig1}).

%%%%%%%%%%%%%%%%%%%%%%%%%%%%%%%%%%%%%%%%%%%%%%%%%%%%%%%%%%%%%

\subsection{Recurrence analysis}
\label{sec2.2}

The tendency to recur to previously visited states is a ubiquitous feature shared by time 
series from many different complex systems. Recurrence plots encode this information in a 
2-dimensional 
binary matrix, indicating a recurrence between two states $\vec{x}_i$ and 
$\vec{x}_j$ at times $i$ and $j$ if the respective states are similar with respect to a given 
norm ${D}(\vec{x}_i, \vec{x}_j)$ \cite{jp1987recurrence}:
\begin{align}
{R}_{ij} \, = \, \left\{
\begin{array}{ll}
1 & \mathrm{if} \ {D}\left(\vec{x}_i, \vec{x}_j\right) \leq \varepsilon \\
0 & \mathrm{if} \ {D}\left(\vec{x}_i, \vec{x}_j\right) > \varepsilon.
\end{array}
\right.
\label{meq4}
\end{align}
The norm ${D}(\vec{x}_i, \vec{x}_j)$ yields a symmetric, real-valued distance 
matrix $\mathbf{D}$ between states at all time instances $i,\,j$. By thresholding $\mathbf{D}$ with the vicinity 
threshold $\varepsilon$, a notion of similar and dissimilar states is implemented and defines the 
recurrence between each pair of states. The underlying idea is based on the Poincaré 
recurrence theorem that states the recurrence of a dynamical system's trajectory 
$\vec{x}(t)$ 
to an $\varepsilon$-neighborhood of any perviously visited state after sufficiently long 
time \cite{poincare1890probleme}. For the main diagonal of the RP, it always holds that 
${R}_{ij} 
\equiv 1$.  If no phase space reconstruction is applied, states 
$\vec{x}_i$ and $\vec{x}_j$ correspond to time series amplitudes ${x}_i$ and ${x}_j$. 
The threshold $\varepsilon$ can be chosen 
based on different data-dependent criteria. In many applications, the recurrence rate is 
fixed to a certain percentage (e.g., $10\,\%$ recurrences \cite{kraemer2018recurrence}) or set to a multiple of the standard 
deviation of the distance matrix $\mathbf{D}$ \cite{schinkel2008}.
The geometric recurrence patterns encoded in a RP can be exploited to distinguish between 
stochastic and deterministic systems \cite{marwan2007recurrence}; while a purely random white 
noise process will result in isolated dots in the recurrence matrix, time series from 
deterministic systems are known to yield diagonal line structures \cite{marwan2007recurrence}. Long diagonal lines are 
characteristic for periodic systems; interrupted diagonal lines indicate chaotic dynamics. 
Recurrence quantification analysis (RQA) which evaluates the statistical properties of a RP has proven a versatile tool for diverse real-world applications, such as time series classification \cite{hirata2021recurrence}, study of causal relations \cite{ramos2017recurrence} or regime shift detection \cite{westerhold2020astronomically}.

Recurrence analysis overcomes some of the flaws of other statistical analysis tools 
when applied to geophysical time series, such as the Lyapunov exponent or correlation dimension 
\cite{kantz1994robust, donner2008nonlinear}. It is less sensitive to noise and can be 
applied to short time series. In combination with the (m)Edit-distance approach, first 
applications demonstrated its ability to detect regime transitions in palaeoclimate proxy 
records \cite{marwan2018regime}. In order to compute a RP for irregularly sampled time series, ${D}(\vec{x}_i, \vec{x}_j)$ in Eq.~(\ref{meq4}) is identified with the 
modified edit distance from Eq.~(\ref{meq1}). In contrast to regular computation of metric 
distances, segments of the time series are required to obtain a distance value between two 
states. Genereally speaking, segment size should be chosen sufficiently small to ensure that 
no aliasing effects arise due to interference between the segment width and the characteristic 
time scale of a time series (e.g., characteristic period of a periodic time series).
For some applications the segments can be chosen such that all are equally sized 
$|\mathcal{S}_{a}| 
= |\mathcal{S}_{b}| = \dots = N$. If this is not possible, the variance of 
segment widths can still be minimized and for each pair of segments with differing widths; 
deletion $\&$ adding operations will contribute to the resulting transformation cost. If time 
series are short, we can allow for an overlap between segments, although caution is advised 
since this introduces a serial dependence in the resulting edit distances of overlapping 
segments and violates the normality assumption used in the estimation of 
$\Lambda_{\mathrm{S}}$.
Here, we focus on the most general case of unequal segment sizes. Apart from cases where 
segment size deviations can hardly be minimized, this is relevant in some real-world 
applications where we are interested in the recurrences between segments that correspond 
to a particular time scale, or where sampling rate is highly non-stationary and selecting a 
constant segment size would result in mixing of distinct time scales. The application to 
palaeoclimate data (Sect.~\ref{sec5}) will illustrate such a case. There, the focus lies on 
the comparison of seasonal sequences in an irregularly sampled proxy time series.

Predictability is a feature of time series that can help to identify and classify different dynamical regimes in the evolution of the studied system. Since the lengths of diagonal lines in a RP reflect the predictability of a system, the number of diagonal lines which exceed a specified minimum line length $l_{\mathrm{min}}$ can be used as a predictiability measure:
\begin{align}
\mathrm{DET} \, = \, \frac{\sum_{l=l_{\mathrm{min}}}^{N}P(l)}{\sum_{l=1}^{N}P(l)}
\label{meq5}
\end{align}
with the number $P(l)$ lines of length $l$.
Determinism (DET) can be linked to the correlation dimension of a dynamical system 
\cite{march2005recurrence} and has successfully been used in diverse empirical analyses 
\cite{ozken2018recurrence,marwan2018regime,westerhold2020astronomically} to detect transitions 
between regimes of varying predictability. We use DET as a recurrence quantifier to test the impact of the sampling-based correction scheme introduced below.

%%%%%%%%%%%%%%%%%%%%%%%%%%%%%%%%%%%%%%%%%%%%%%%%%%%%%%%%%
%%%%%%%%%%%%%%%%%%%%%%%%%%%%%%%%%%%%%%%%%%%%%%%%%%%%%%%%%
%%%%%%%%%%%%%%%%%%%%%%%%%%%%%%%%%%%%%%%%%%%%%%%%%%%%%%%%%
%%%%%%%%%%%%%%%%%%%%%%%%%%%%%%%%%%%%%%%%%%%%%%%%%%%%%%%%%

\section{Segment size dependence}
\label{sec3}
%\mathbf{E}[k; \lambda_1, \lambda_2] \ = \ 2\mathrm{e}^{-\lambda_1-\lambda_2}\left( \lambda_2I_0(2\sqrt{\lambda_1\lambda_2}) + \sqrt{\lambda_1\lambda_2}I_1(2\sqrt{\lambda_1\lambda_2}) \right) \, + \, (\lambda_2 - \lambda_1)\left(1-2Q(\sqrt{2\lambda_1},\sqrt{2\lambda_2})\right)
Finite-sample effects are known to entail statistical biases in various time series analysis methods. Linear or spline interpolation is often employed as a pre-processing technique to enable the application of standard time series analysis tools to irregularly sampled time series. Interpolation techniques do not account for basic finite-sample biases. 
For instance, statistical location and scale measures (such as the median or volatility 
indicators) are known to be biased for small sample sizes \cite{williams2011finite, 
park2019investigation}. Given two segments $\mathcal{S}_{a},\, \mathcal{S}_{b}$ with 
$|\mathcal{S}_{a}| 
\gg |\mathcal{S}_{b}|$, estimating their variance (e.g., as a volatility 
indicator or in order to compute a continuous wavelet spectrum) can result in underestimation of the variance for the shorter segment. 
Similarly, persistence estimators are generally biased due to finite-sample effects, even for 
Markovian stationary stochastic processes \cite{trenberth1984some}. Whenever a sliding-window 
analysis for nonstationary, irregularly sampled time series is carried out, variations in the 
sampling rate will inevitably result in a mixture between the actual variability of the statistical 
indicator and purely sampling-related variations. As interpolation techniques are usually 
limited to resampling values such that sampling intervals are equal, this effect is not 
compensated. Similar intricacies need to be considered in short time series, e.g., 
when computing correlations between multiple time series (of varying length) \cite{goswami2017inferring}.

\begin{figure*}[]
\centering
\includegraphics[width=.9\textwidth]{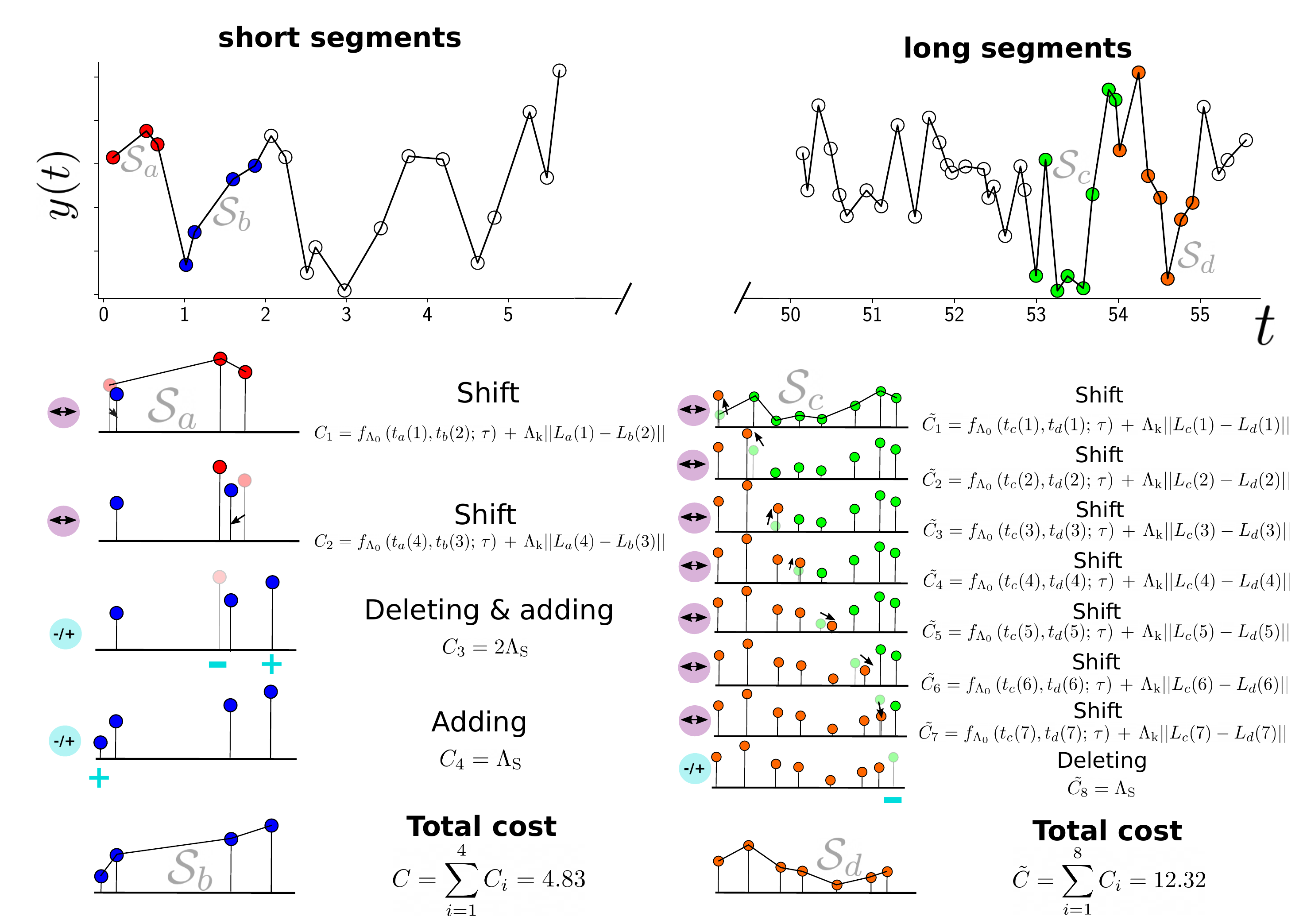}\\
\caption{Schematic illustration of how irregularly sampled segments of varying lengths are 
transformed with the (m)Edit-distance method. Two exemplary pairs of segments 
$\mathcal{S}_a,\, 
\mathcal{S}_b$ (\textbf{A}: red, blue) and $\mathcal{S}_c,\, \mathcal{S}_d$ (\textbf{B}: green, 
orange) of an irregularly sampled synthetic AR(1)-time series are displayed. Each row shows an 
operation applied to the respective segment (shift: purple, deletion/adding: cyan). Final 
costs $C$ and $\tilde{C}$ result from a specific choice of cost parameters as described in 
Sect.~\ref{sec2.1}. Please note that the higher total cost in \textbf{B} showcase the dependence on segment length.}
\label{fig1}
\end{figure*}

While not designed to compensate such effects, the (m)Edit-distance methodology does not introduce any known additional biases. The computation of 
transformation costs is demonstrated with two exemplary pairs of segments $\mathcal{S}_a,\, 
\mathcal{S}_b$ and $\mathcal{S}_c,\, \mathcal{S}_d$ (Fig.\ref{fig1}). The 
segments $\mathcal{S}_a,\, \mathcal{S}_b$ all display distinct operations for transforming a 
segment into another: in the first step, a shift of amplitude and time are applied to transform the time instance 
$t_a(1)$ and amplitude $L_a(1)$ of the first segment into time instance $t_b(2)$ 
and amplitude $L_b(2)$ of the second segment. The cost $C_1$ associated with this operation is 
the sum of shifting both time and amplitude. After shifting the third value of $\mathcal{S}_a$ 
to match the third value of $\mathcal{S}_b$, both a deletion and an adding operation are 
performed in step 3 with twice the cost $\Lambda_{\mathrm{S}}$ for a adding/deleting 
operation. The same transformation could have been achieved with an additional shifting 
operation. The prefered operation is determined by the particular choice of cost parameters. 
As $|\mathcal{S}_a|=3$ and $|\mathcal{S}_b|=4$, the first value of $\mathcal{S}_b$ is added in 
step 4. The resulting cost is the sum of all costs for each step. While different 
transformation paths are possible, the algorithmic implementation ensures that $C$ is 
minimized with respect to all possible combinations. Another example is displayed in the right 
column of Fig.\ref{fig1}. The setup differs in that the indicated segments 
$\mathcal{S}_c,\, 
\mathcal{S}_d$ are longer than $\mathcal{S}_a,\, \mathcal{S}_b$ ($|\mathcal{S}_c|=8,\, 
|\mathcal{S}_d|=7$). 
Despite a similar set of transformations, the resulting costs 
$\tilde{C}$ 
are significantly higher for the exemplary choice of parameters.

A systematic derivation of transformation costs on segment size/sampling rate for exponentially distributed sampling intervals is given in appendix A.
Note that the identified effect is not due 
to an immanent misconception in the edit distance computation. It solely arises from the 
fact that the edit distance is applied in a setting where the time axis is not only irriations in its sampling rate.
In particular, abrupt transitions in the sampling rate between a time period $T_1$ with low 
sampling rate $\lambda_1$ and $T_2$ with high sampling rate $\lambda_2$ will imprint a 
non-trivial 
$\lambda_1,\,\lambda_2$-dependence on the transformation cost  
$D(\mathcal{S}_a\,\mathcal{S}_b)$ 
between any two segments.
In a recurrence analysis of time series, the focus lies on the similarity of states based on egular but undergoes 
significant var
the amplitudes of the time series. Hence, we argue that the identified dependencies counteract 
the goal of recurrence analysis of irregularly sampled time series and thus need to be 
corrected such that recurrence quantification measures reflect the dynamical behaviour of the 
underlying system rather than mere shifts in the sampling rate.

We numerically examine the dependence of transformation costs between segments $\mathcal{S}_{a},\, \mathcal{S}_{b}$ on their sizes $N_a, N_b$ for simple synthetic time series.
We test irregularly sampled time series from three different model systems: uncorrelated 
uniform noise, an AR(1)-process ($\tau=5$), and a sinusoidal ($\nu =\left.1 \middle/ 25\right.$) 
with superimposed low-amplitude white noise. Segments of specified sizes from each of these 
systems are drawn to compute segment size-specific costs.
\FloatBarrier
Irregular time axes are generated from a $\gamma(\Delta; k,\Theta)$-distribution with scale 
$\Theta$ and shape $k = \sqrt{\left. 2 \middle/ \Gamma \right.}\,$, where $\Gamma$ denotes the 
skewness of the distribution.
This choice is motivated by the observation that sampling 
intervals in palaeoclimate proxy time series are often $\gamma$- rather than 
exponentially-distributed. 
For each system, we generate a `superpopulation' $(K=100)$ of time series and 
time axes. Fixing a different skewness $\Gamma$ of the $\gamma\,$-distribution of each 
of the time axes between $\Gamma\in[1,8]$ ensures that for $T=10,000$, segment sizes range 
between $N\in[1, 20]$. The modified edit-distance is used, Eq.~(\ref{meq1}) and deletions are included as a competing operation to shifting.
The optimal $\Lambda_{\mathrm{S}}$ is estimated for each system according to the procedure 
outlined in Sect.~\ref{sec2.1}: the KS-statistic is minimized for each systems, yielding
$\Lambda^{\mathsmaller{(\mathrm{unif})}}_{\mathrm{S}} = 1.5,\, \Lambda^{\mathsmaller{(\mathrm{AR1})}}_{\mathrm{S}} = 1.5,\, \Lambda^{\mathsmaller{(\mathrm{sin})}}_{\mathrm{S}} = 3.5$.

\begin{figure*}[]
\centering
\begin{subfigure}[t]{1\textwidth}
\centering
\includegraphics[width=.32\textwidth]{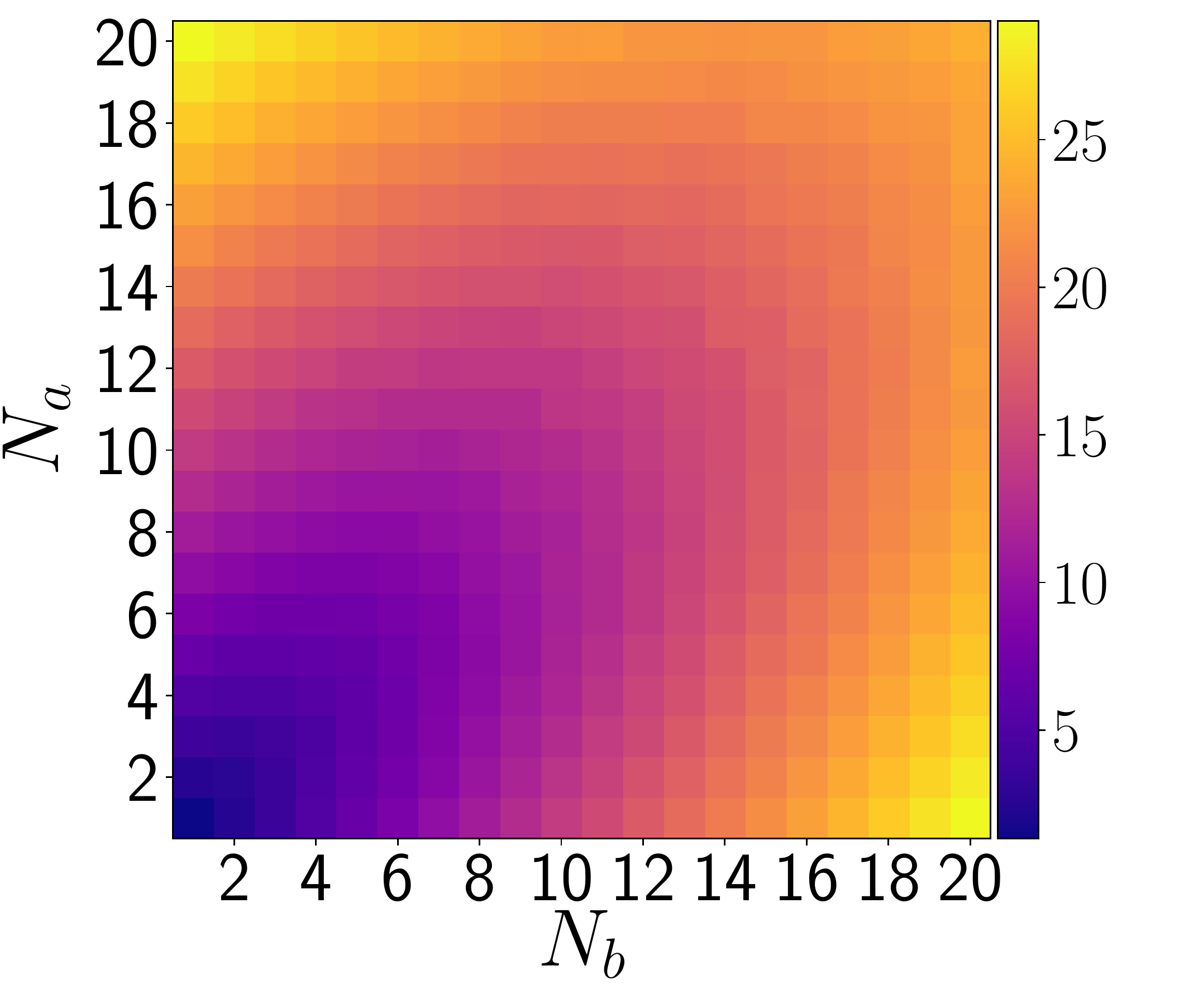}
\includegraphics[width=.32\textwidth]{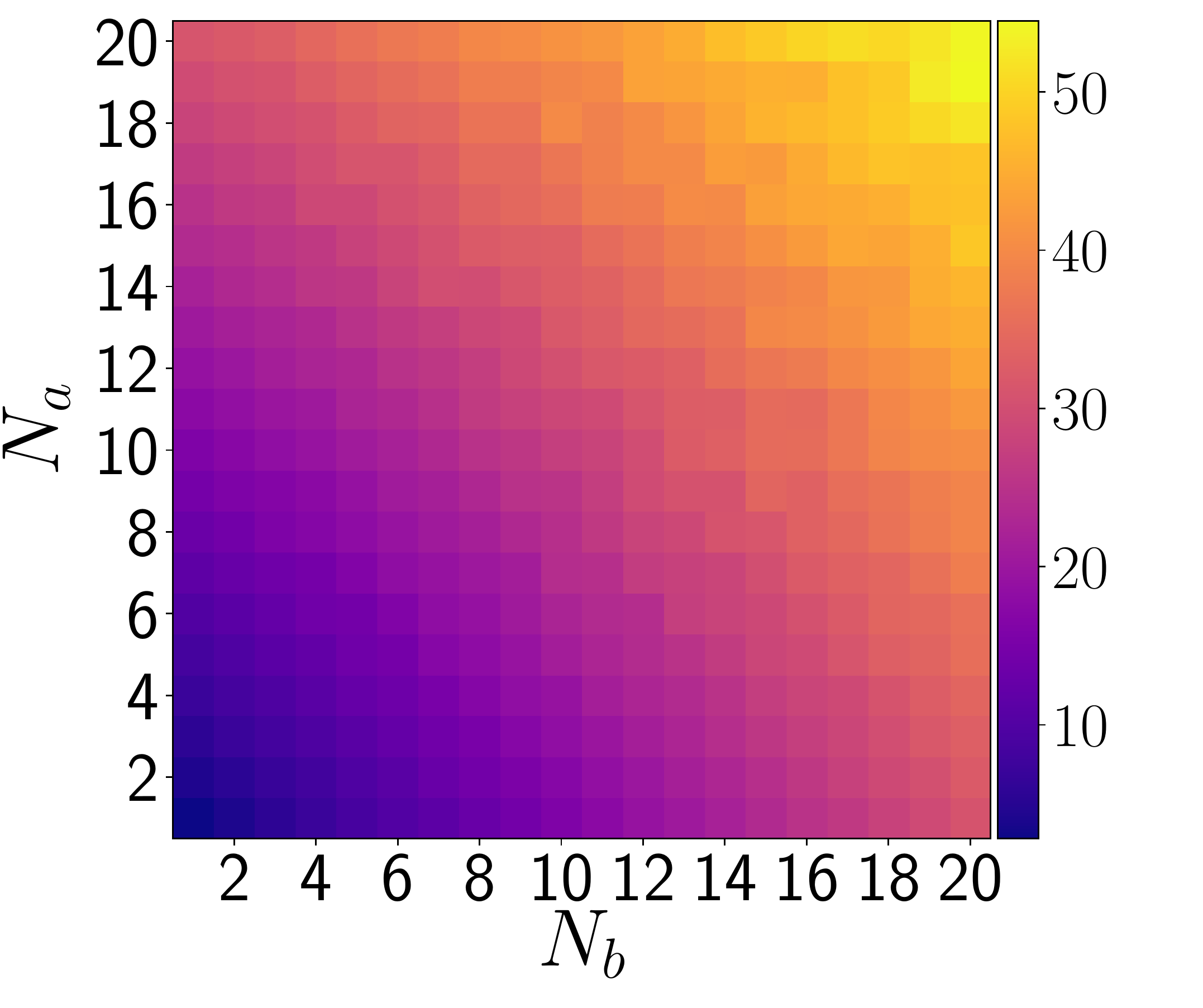}
\includegraphics[width=.32\textwidth]{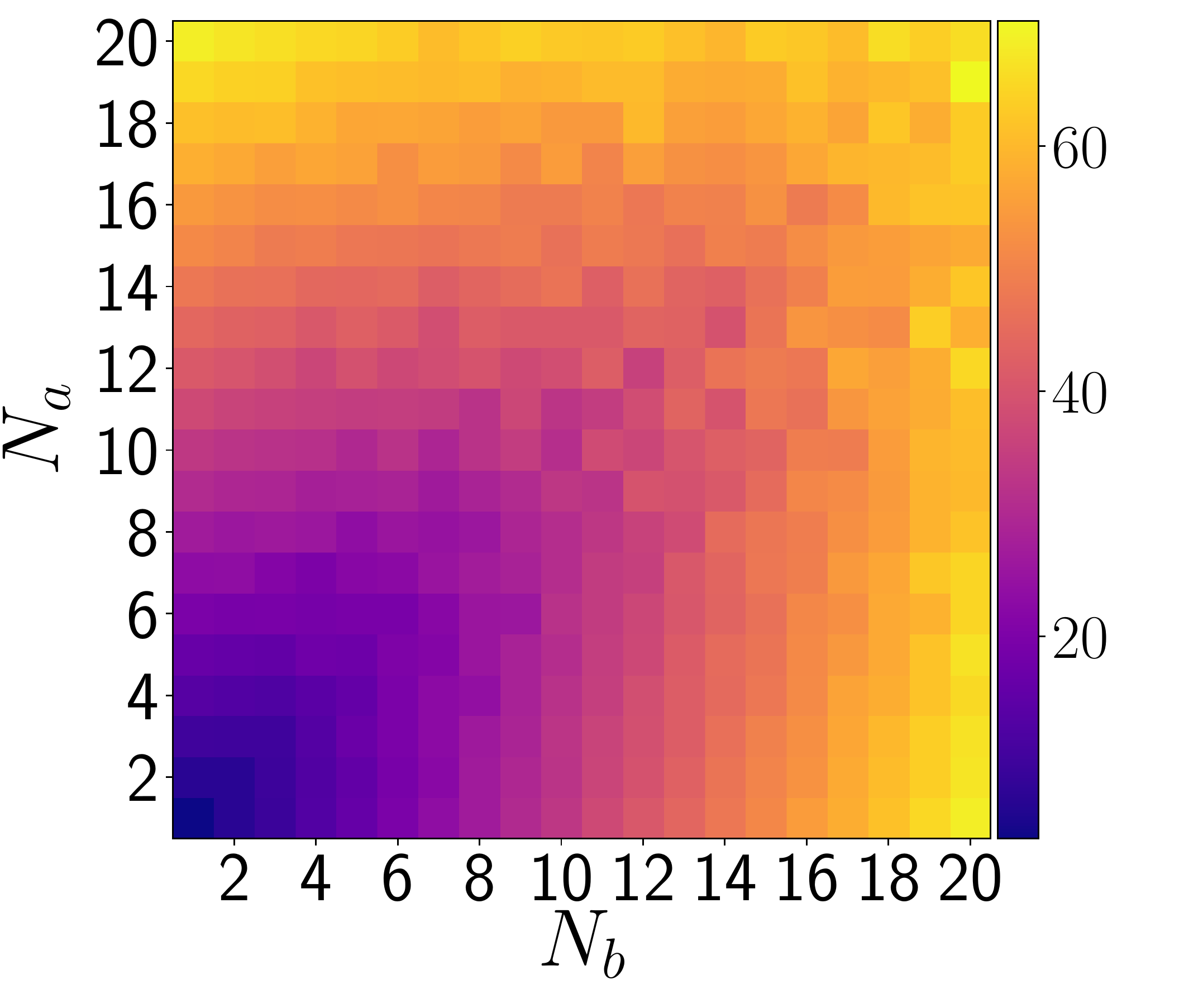}
\caption{Cost matrices ${{C}}(N_a,N_b)$ (including basic deletions, total costs).}
\label{fig2a}
\end{subfigure}
\begin{subfigure}[t]{1\textwidth}
\centering
\includegraphics[width=.32\textwidth]{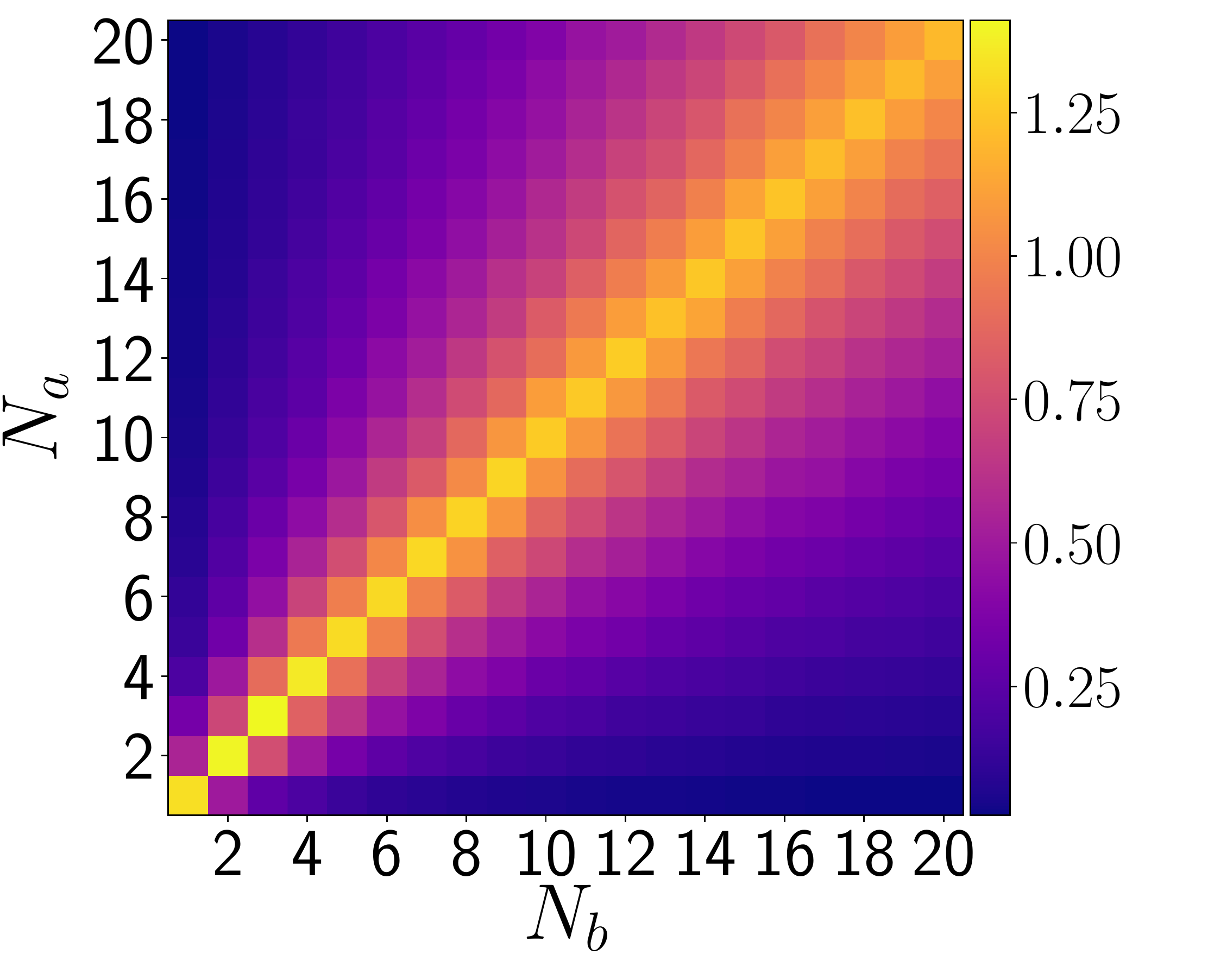}
\includegraphics[width=.32\textwidth]{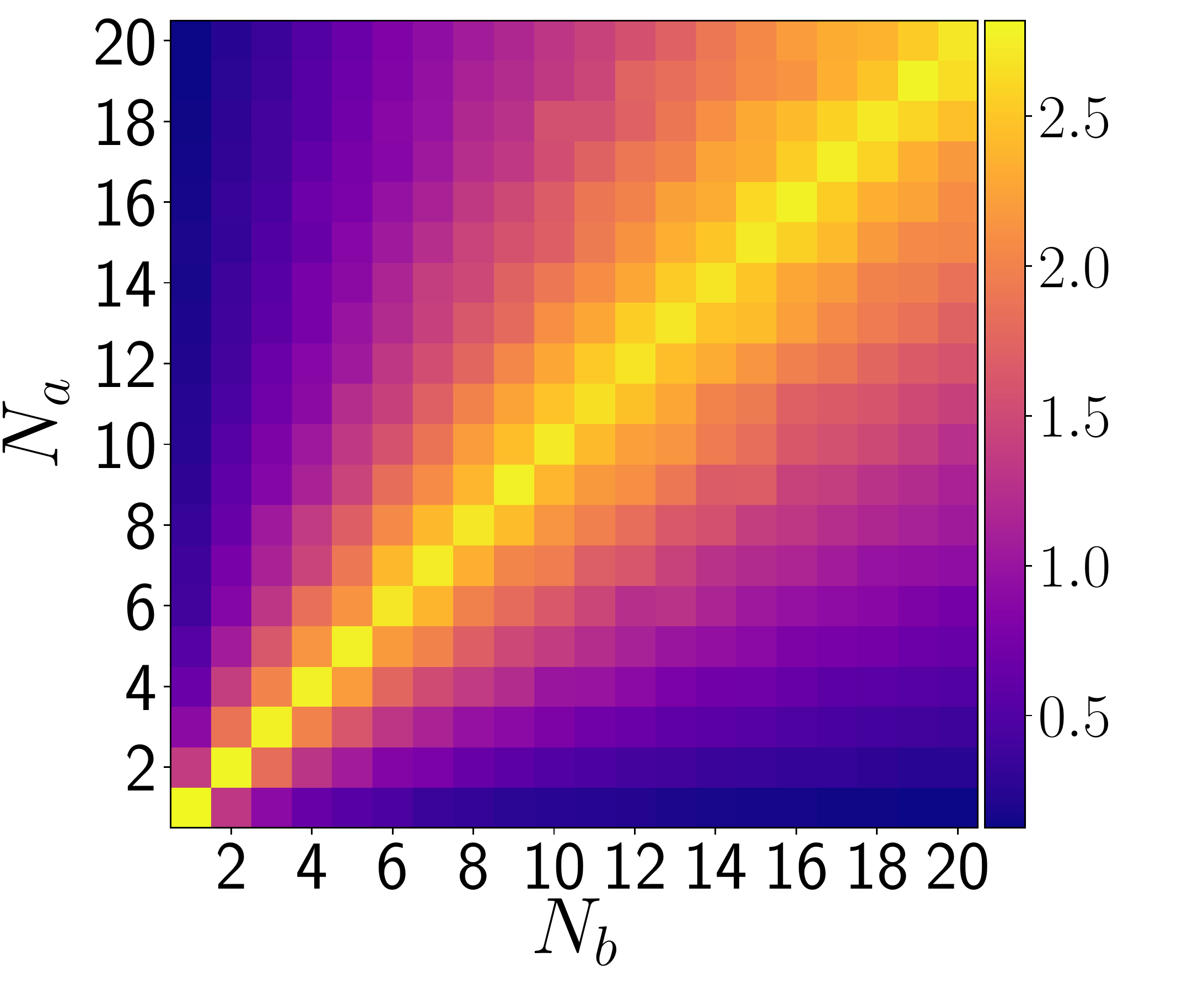}
\includegraphics[width=.32\textwidth]{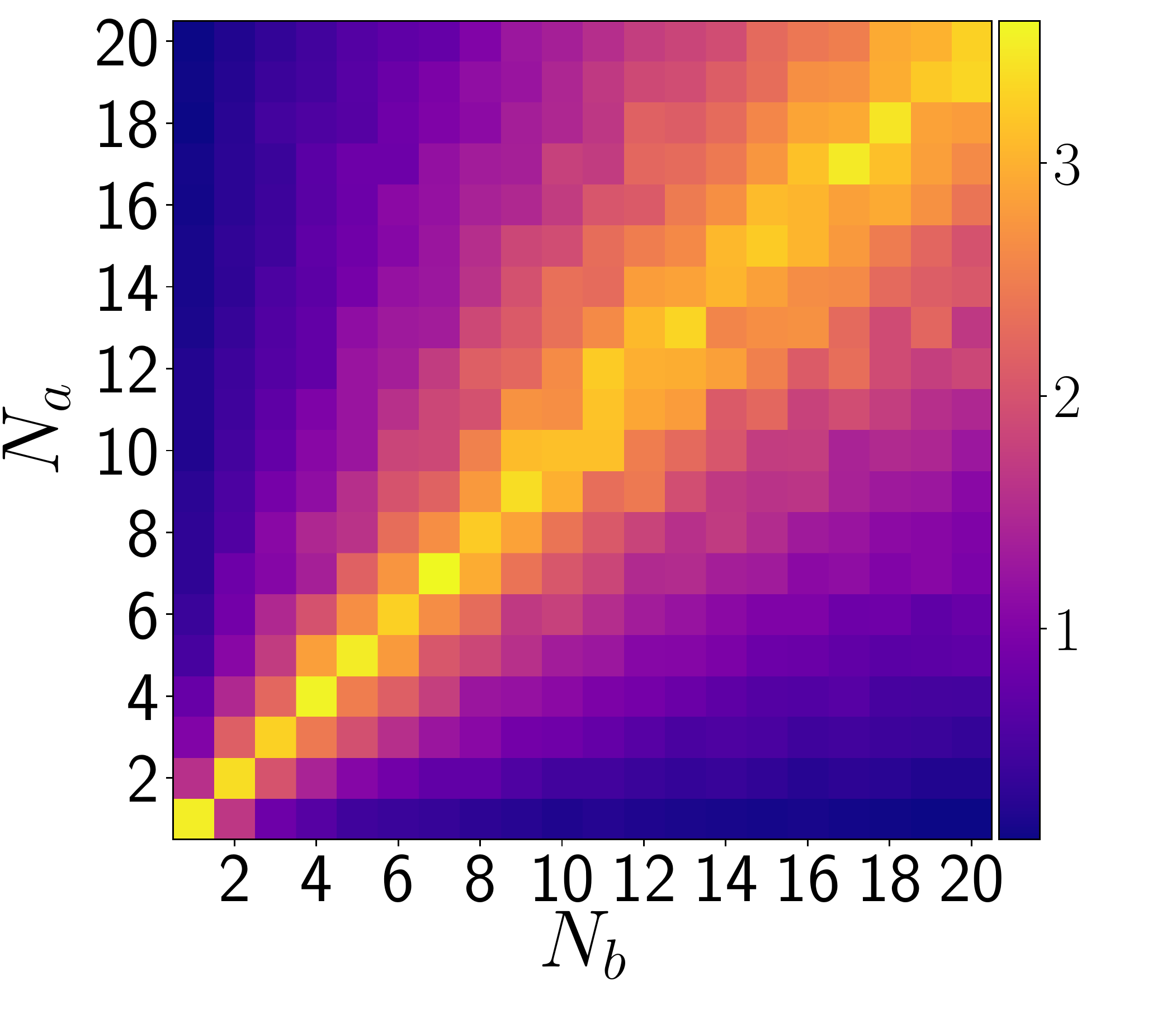}
\caption{Cost matrices ${{\tilde{C}}}_{\mathrm{shift}}(N_a,N_b)$ (excluding basic deletions, average costs per operation).}
\label{fig2b}
\end{subfigure}
\caption{Cost matrices ${{C}}(N_a,N_b)$ for the transformation of segments with different lengths, including basic deletions (a) and excluding basic deletions (b). Costs are shown for uncorrelated uniform-distributed noise (left), an AR(1)-process (center) and a sinusoidal with superimposed white noise (right). Sampling intervals are $\gamma\,$-distributed.}
\label{fig2}
\end{figure*}

Figure \ref{fig2a} displays the obtained transformation costs in the cost matrices 
${{C}}(N_a,N_b)$ 
and ${{\tilde{C}}}_{\mathrm{shift}}(N_a,N_b)$ 
after averaging over $K=100$ different realizations. Regardless of the irregularity of the 
time axis and the respective system, a tendency of increasing total costs for larger segment 
sizes is observed (upper row). For the AR(1)-system, this increase is slower for fixed 
$N_b$ and increasing $N_a$ composed to the uncorrelated noise and the sinusoidal examples. More generally, 
the rate of increase differs between the considered systems but follows the same trend.
In total, $|N_a-N_b|$ `basic deletions' (or adding operations) need to be carried out for each pair of segments with $N_a\neq N_b$.
If costs for these basic deletions are subtracted and computed per shifting step, a similar 
dependency on $N_a,\,N_b$ as observed in Fig.~\ref{figA1}c for the more simple case can be 
observed in the cost matrices ${{\tilde{C}}}_{\mathrm{shift}}(N_a,N_b)$ in 
Fig.~\ref{fig2b}\,: the cost of an average shift from a segment with $N=N_a$ increases towards 
$N_b=N_a$ and decays if segment size increases further.
Consequently, the leading effect results from the basic deletions that are directly linked to the difference in segment sizes $|N_b-N_a|$. Yet, transformation costs still depend on segment size even after aligning both segment sizes by means of basic deletions; this effect likely results from having a higher probability of finding closely spaced values on the time axis as the sampling rate of one segment increases, yielding an increasing trend for average costs per operation (in 
fig.~\ref{fig2b}).
%%%%%%%%%%%%%%%%%%%%%%%%%%%%%%%%%%%%%%%%%%%%%%%%%%%%%%%%%
%%%%%%%%%%%%%%%%%%%%%%%%%%%%%%%%%%%%%%%%%%%%%%%%%%%%%%%%%
\section{Sampling rate constrained surrogates}
\label{sec4}
Irregularly sampled time series with constant sampling rate can be studied with the 
(m)Edit-distance to obtain dissimilarity estimates between different time series 
segments. The resulting distance matrix can be used to perform a recurrence analysis. 
Moreover, other analysis techniques such as complex networks, clustering, or correlation analysis are based on (dis)similarity measures and could use the (m)Edit-distance as a metric to account for irregular sampling or to characterize event-like 
data. In section \ref{sec3} we showed that in case of a non-constant sampling rate, an 
estimation of the (m)Edit-distance matrix is biased by significant differences in the 
segment sizes.

In the following, we propose a numerical correction-technique for recurrence analysis. We generate an ensemble of time series and time axis surrogates that reproduces the sampling properties of the real irregularly sampled time series. This surrogate ensemble is used for 
bias-correction of recurrence quantification measures, exemplified by the determinism DET.
\subsection{Constrained randomization}
\label{sec4.1}
When studying a system's dynamics with time series analysis tools, a null-hypothesis 
is formulated which can be be tested. In case of recurrence analysis, 
this hypothesis could for example be non-stationarity of a dynamical property of the 
system (predictability, serial/cross-dependence, \ldots) expressed by a particular 
recurrence quantification measure. In the used example, the 
null-hypothesis 
tests whether the observed dynamics could be solely caused by variations in the sampling rate. 

\begin{figure*}[htbp]
\centering
\includegraphics[width=1\textwidth]{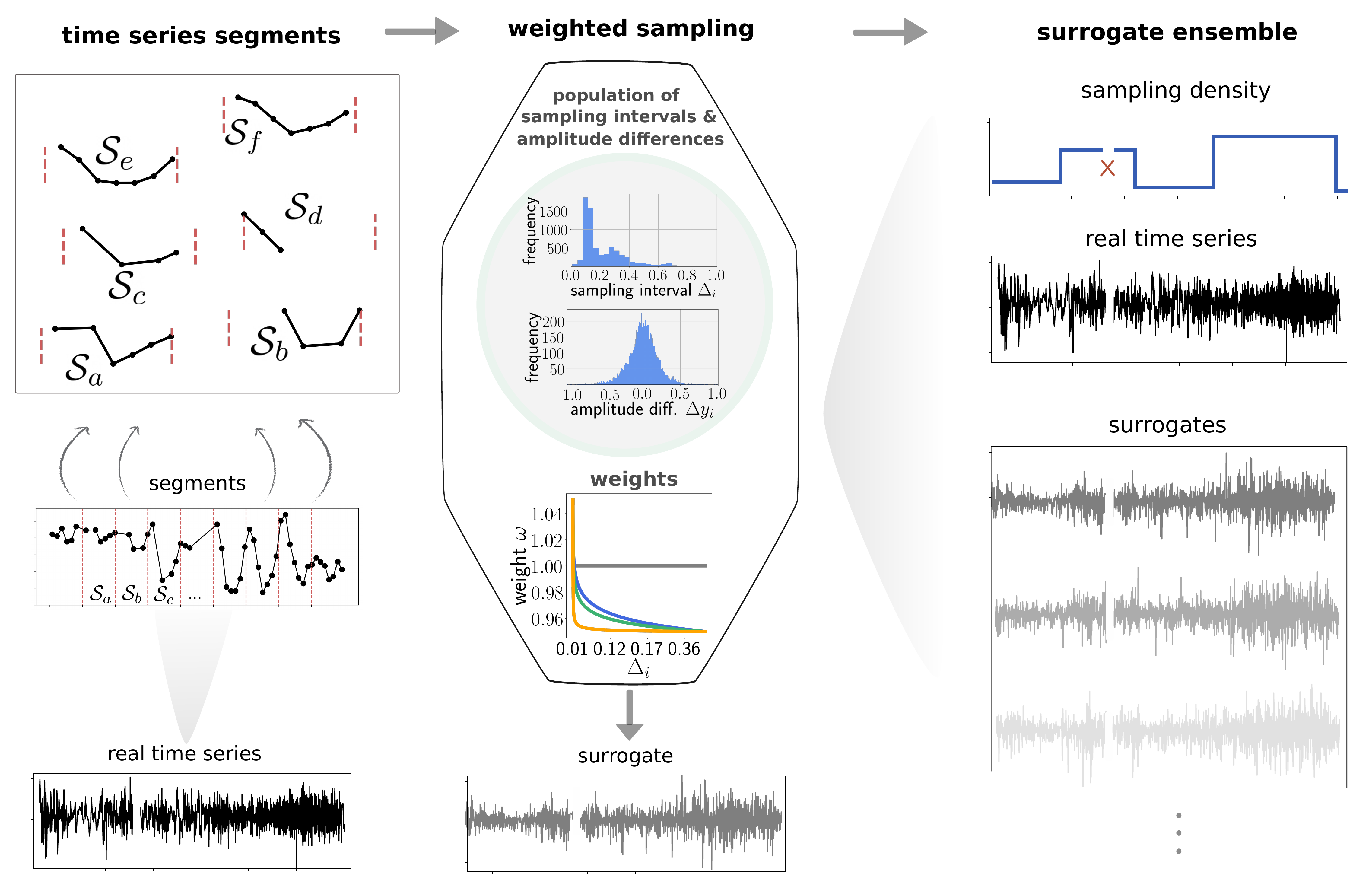}
\caption{Schematic illustration of the constrained randomization procedure that generates SRC-surrogates for an exemplary irregularly sampled time series with non-stationary sampling rate. The left column shows the segmentation of the time series into segments of constant time period $s$ but of variable size $N_i$. The center column illustrates the weighted sampling of sampling intervals and amplitude differences. Each sampling interval is assigned a $\beta\,$-distributed weight whereby the $\alpha$-parameter of the weight distribution is increased with each $l$-th failed iteration to favor short sampling intervals. The resulting surrogates preserve the empirical distributions and segment sizes. Since amplitude differences are sampled jointly with the respective sampling intervals, increased volatility simply due to a higher local 
sampling rate is reproduced by the SRC-surrogates.}
\label{fig3}
\end{figure*}

Parametric hypothesis testing for time series analysis often poses severe constraints 
on the statistical properties of the underlying probability distribution, e.g., 
normality. Surrogate tests represent a non-parametric and flexible method to test for 
a range of properties in a system, including nonlinearity or periodicity, among 
others  \cite{schreiber2000surrogate}. Time series surrogates are altered copies of a 
real, underlying time series that only preserve a specified set of properties of the 
real time series. The general technique to generate surrogate realizations of a time 
series is constrained randomization. After defining a set of constraints that state 
which properties of the real time series should be preserved, the time series is 
randomized such that these constraints are still fullfilled. Here, randomization will be carried out on the sampling interval $\Delta_i$ with 
the constraint that for each segment $\mathcal{S}_i$ of the real time series, segment 
size $N_i$ is preserved. This is achieved by drawing sampling intervals $\Delta_i$ 
(with replacement) from the empirical sampling interval distribution $p(\Delta, 
\lambda (t))$. For a given segment $S_i$ with size $N_i$, $N_i$ 
sampling intervals are drawn from $p(\Delta, \lambda (t))$ and cumulated to generate a 
surrogate realization of the particular time axis segment: 
\begin{align}
\tilde{t}_{\mathcal{S}_i}^{(0)} = t_{\mathcal{S}_i}^{(0)} \, , \quad  \tilde{t}_{\mathcal{S}_i}^{(j+1)} = \tilde{t}_{\mathcal{S}_i}^{(0)} + \sum_{m=0}^j \Delta_{i}^{(m)}
\label{meq6}
\end{align}
Let $w$ be the time period covered by each segment. For any randomly sampled set of 
sampling intervals, the constraint of preserved segment size requires that
\begin{equation}\label{condition_t}
\tilde{t}_{\mathcal{S}_i}^{(N_i)} \stackrel{\text{!}}{\leq} w, 
\end{equation}
otherwise the random sampling of sampling intervals $\Delta_i$ has to be repeated. 
If the distribution of segment sizes is short-tailed, i.e., no segments with size 
$N\gg \mathbb{E}[k]$ exist, this simple randomization procedure converges rapidly for 
each segment. If segments of relatively large size are present -- which is likely the 
case for non-stationary sampling rates -- only a small subset of sampling intervals from 
the left tail of $p(\Delta, \lambda (t))$ will fulfill the condition 
(\ref{condition_t}). In order to ensure convergence of the algorithm for large 
segments, a weight-function can be introduced for all sampling intervals to increase 
the likelihood of drawing short sampling intervals when a segment with large size is 
generated. We suggest the use of $\beta$-distributed weights $\omega$:
\begin{align}
\omega(X;\,\alpha, \beta) \, = \, \frac{1}{B(\alpha, \beta)}x^{\alpha-1}(1-x)^{\beta-1}
\label{meq7}
\end{align}
with the $\beta$-function $B(\alpha, \beta)$. This choice is motivated by the fact 
that for $\alpha = \beta = 1$, $\omega(X;\,\alpha, \beta)$ becomes a uniform 
distribution. In our application, we choose $\alpha = \beta = 1$ when the first 
iteration of sampling $N_i$ sampling intervals $\Delta_i$ is carried out. The 
population of sampling intervals is ordered from shortest to largest and each 
$x_i\leftrightarrow \Delta_i$ is assigned a $\beta$-distributed weight $\omega_i$, 
i.e., for the first iteration, every sampling interval is drawn with equal 
probability. If the iteration fails $\left(\tilde{t}_{\mathcal{S}_i}^{(N_i)} > 
w\right)$, $\alpha$ is increased by a small number $\Delta\alpha$, reshaping the 
beta-distribution 
and increasing the probability of drawing small sampling intervals. Thus, 
we perform a weighted sampling from the empirical distribution $p(\Delta, \lambda 
(t))$ 
of sampling intervals with $\beta$-distributed weights. In the $l$-th 
iteration, we use $\omega(X;\,\alpha_l, \beta=1),\, \alpha_l = 1 + l\Delta\alpha$ as 
the weight function for each segment. Finally, we can identify an amplitude difference 
$\Delta y_i$ of the time series with each sampling interval $\Delta_i$. This 
correspondence is exploited by also drawing the respective amplitude difference for 
each drawn sampling interval. After the procedure is finalized and a surrogate has 
been generated, amplitude differences are cumulated, that yields both a time axis and 
time series surrogate. Both are denoted as sampling rate constrained 
surrogate (SRC-surrogates).
The full randomization procedure thus preserves segment sizes in the correct temporal 
order and by definition approximately reproduces the distribution of amplitude differences and 
sampling intervals.
\FloatBarrier
 It also preserves the correspondance between sampling 
intervals and amplitude differences, ensuring that if periods with high local sampling 
rate entail larger variance/strong amplitude changes in a real time series, this 
property is also included in the SRC-surrogates. The full procedure is outlined in 
Fig.~\ref{fig3} for an exemplary time series. 
Other randomization schemes are conceivable, e.g., varying the sampling weights after 
drawing each single sampling interval based on the size of the latter, or stratified 
randomization, i.e., performing the randomization differently for strata that correspond to the 
different segment sizes. However, the proposed scheme has proven to be effective within the scope of this work.

With the presented scheme of generating SRC-surrogates, an ensemble of surrogates can be 
generated and (m)Edit-distance matrices $\mathbf{D}$ computed for each SRC-surrogate. 
Any measure that is based on $\mathbf{D}$ can consequently be computed for each surrogate 
separately, yielding a distribution that can be used for testing the null-hypothesis 
formulated above based on the desired $\alpha$-confidence level.
%%%%%%%%%%%%%%%%%%%%%%%%%%%%%%%%%%%%%%%%%%%%%%%%%%%%%%%%%
%%%%%%%%%%%%%%%%%%%%%%%%%%%%%%%%%%%%%%%%%%%%%%%%%%%%%%%%%
\subsection{Recurrence analysis of an AR(1)-process}
\label{sec4.2}

In the example below, the proposed correction scheme is applied to an irregularly sampled AR(1)-process (Fig.~\ref{fig5}a).
We consider an autocorrelation increasing with time, visible by autocorrelation time $\tau$ (Fig.~\ref{fig5}b).
A recurrence analysis is used to characterize the predictability of the time series in a sliding window analysis. 
Predictability is computed by means of determinism, DET, as defined in Eq.~(\ref{meq5}).
In particular, we study how an abrupt shift of the sampling rate (represented by the skewness of $\gamma$-distributed 
sampling intervals) affects DET and if a continuous increase of predictability can be recovered despite this shift by using 
the proposed SRC-surrogate method. The shift appears at $t=1250$ (visible by variation of the segment size, Fig.~\ref{fig5}b).
\FloatBarrier
We expect DET to reproduce the linear increase in autocorrelation time, because increased serial dependence implies longer 
and more diagonal lines in the RP.
For the computation of the (m)Edit-distance measure, segments are picked such that each covers a constant time interval of 
$w = 1$ which could correspond to a year in a real-world application. 200 SRC-surrogates are generated (see appendix B) with $\alpha_0=1, \beta=1$ and a step size for the shape parameter 
$\alpha$ of the beta-distribution $\Delta\alpha=0.15$.
We set an upper limit of $N_{\mathrm{it}}^{(\mathrm{max})}=1000$ for 
the number of iterations in the generation of each segment which is never exceeded in the performed simulations. 
The deletion/adding cost parameter $\Lambda_S$ is estimated separately for the real time series and the surrogate realizations, yielding $\Lambda_S^{(\mathrm{real})} = 5.3$ and $\Lambda_S^{(\mathrm{SRC})} = 2.6$. Recurrence plots are 
computed on sliding windows of size $s=200\Delta$ with $75\%$ overlap (time series length: $T=5000$).
We fix a recurrence rate of $15\%$ and do not apply any time-delay embedding.
For each window, two DET values are obtained (Fig.~\ref{fig5}c): the DET value of the real time series and the 
$\alpha(=95\%)$-confidence 
level of DET values calculated from the SRC-surrogate ensemble. The DET measure indicates a 
spurious transition of predictability induced by the abrupt shift in sampling rate (Fig.~\ref{fig5}c, grey shading). Both the real time series 
and the surrogate ensemble indicate this shift, demonstrating that the proposed SRC-surrogates effectively reproduce the 
sampling bias. 
\begin{figure*}[]
\centering
\includegraphics[width=.7\textwidth]{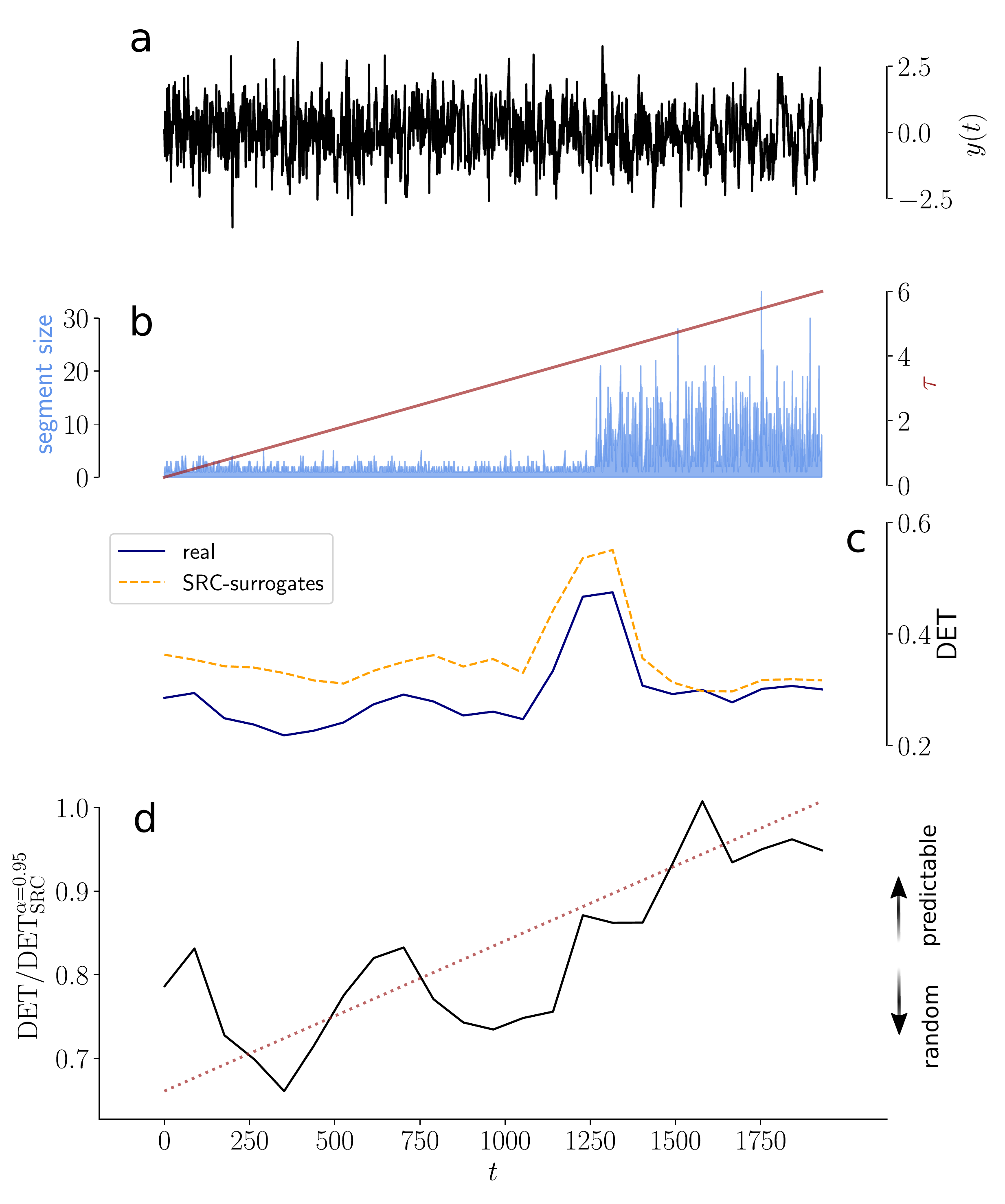}
\caption{Application of SRC-surrogate correction method to \textbf{(a)} an irregularly sampled AR(1)-process with 
\textbf{(b)} non-stationary sampling rate (blue) and linearly increasing autocorrelation time (red). Gray shading indicates 
the abrupt shift in sampling rate. \textbf{(c)} A sliding window RQA using determinism (DET) as a predictability measure is 
carried out. Real DET values are displayed in dark blue, the $95\%$-confidence level computed from 200 SRC-surrogates is 
shown in yellow. \textbf{(d)} The ratio $\mathrm{DET}_{\mathrm{real}}$ by $\mathrm{DET}_{\mathrm{surr}}$ provides a 
sampling-bias corrected predictability measure that reproduces the linear increase in serial dependence.}
\label{fig4}
\end{figure*}
%\FloatBarrier
%
The SRC-based correction is applied to DET values by dividing the real 
DET-series 
by the $95\%$-confidence level for each window (Fig.~\ref{fig5}d). The resulting predictability estimates reproduce the expected linear 
increase in serial dependence whilst eliminating the spurious shift due to the jump of the sampling rate.
\section{Real-world application: rainfall seasonality in the central Pacific}
\label{sec5}
Many real-world proxy time series are characterized by irregular sampling or missing data and stationarity of the underlying process that controls the 
sampling rate cannot be guaranteed. This perspective even goes beyond uneven time axes as for some systems, it might be 
desirable to apply an adaptive windowing in order to obtain segments with segment sizes depending on specific parameters of 
the system. For instance, when analyzing cardiac time series it might be reasonable to choose the segment size adaptively to capture one heart-beat cycle within each segment. The length of every cycle is controlled by a variety of other physical, non-stationary parameters.
Below, we focus on an irregularly sampled palaeoclimate proxy time series with a non-stationary temporal sampling rate.
We demonstrate the effectiveness of the proposed approach by carrying out a sliding window recurrence analysis.

The palaeoclimate record analysed here is a seasonally resolved stalagmite proxy record from Niue Island in the southwestern Pacific (19$^{\circ}$S, 169$^{\circ}$W). It covers 1000 years in the mid-Holocene (6.4-5.4 thousand years before present (ka BP)). 
Niue island has a tropical climate, receiving an average of 2000 mm of precipitation annually with a pronounced wet season from November to April. Rainfall is most strongly controlled by seasonal displacement of the South Pacific Convergence Zone but also reacts sensitively to atmospheric circulation changes associated with the El Niño-Southern Oscillation. 
Here, we analyse seasonal rainfall variability on Niue recorded in greyscale changes that arise from crystallographic variations caused by changes in the stalagmite growth rate (Fig.~\ref{fig6}). Greyscale values are obtained from high resolution scans of the stalagmite surface along its growth axis subsequently extracted with ImageJ \cite{abramoff2004image}. 
During the dry season, low drip rates promote the deposition of layers with compact and dark crystals, yielding low greyscale values. In the wet season, the drip rates are higher and crystal growth is enhanced as dissolved inorganic carbon is supplied to a greater extent. (see Fig.~\ref{fig6}). The inferred link between dark layers and dry season is supported by earlier studies \cite{aharon2006caves} \citep{CINTHYA}.

Prior to the recurrence analysis of the greyscale time series, we 
subtracted a centennial-scale trend using a Gaussian kernel filter in order tp focus on the high-frequency variability in the record (Fig.~\ref{fig6}a, black line). Next, we downsampled the time series uniformly by only storing every 10$^\text{th}$ value due to computational constraints. 
This downsampling does not alter the relative changes in the sampling rate (Fig.~\ref{fig6}b). The number of samples per year (i.e., the segment 
size) undergoes an abrupt shift at $\approx\,$6.15~ka~BP. The period with the highest average segment size ($\approx\,$ 6.4 to 6.15~ka BP) coincides with the wettest period covered by the record, indicated by high greyscale values. 
This suggests that during this wet period, stalagmite growth was enhanced which resulted in thicker crystal laminae and a higher number of samples per layer. This observation reflects the complex nature of irregular sampling of palaeoclimate-proxy data. If spatial 
sampling on the stalagmite is performed such that the number of samples is as high as possible, it will inevitably be linked 
to its growth rate and thus to other environmental parameters and their non-stationary characteristics.
\begin{figure}[htbp]
\centering
\includegraphics[width=.7\textwidth]{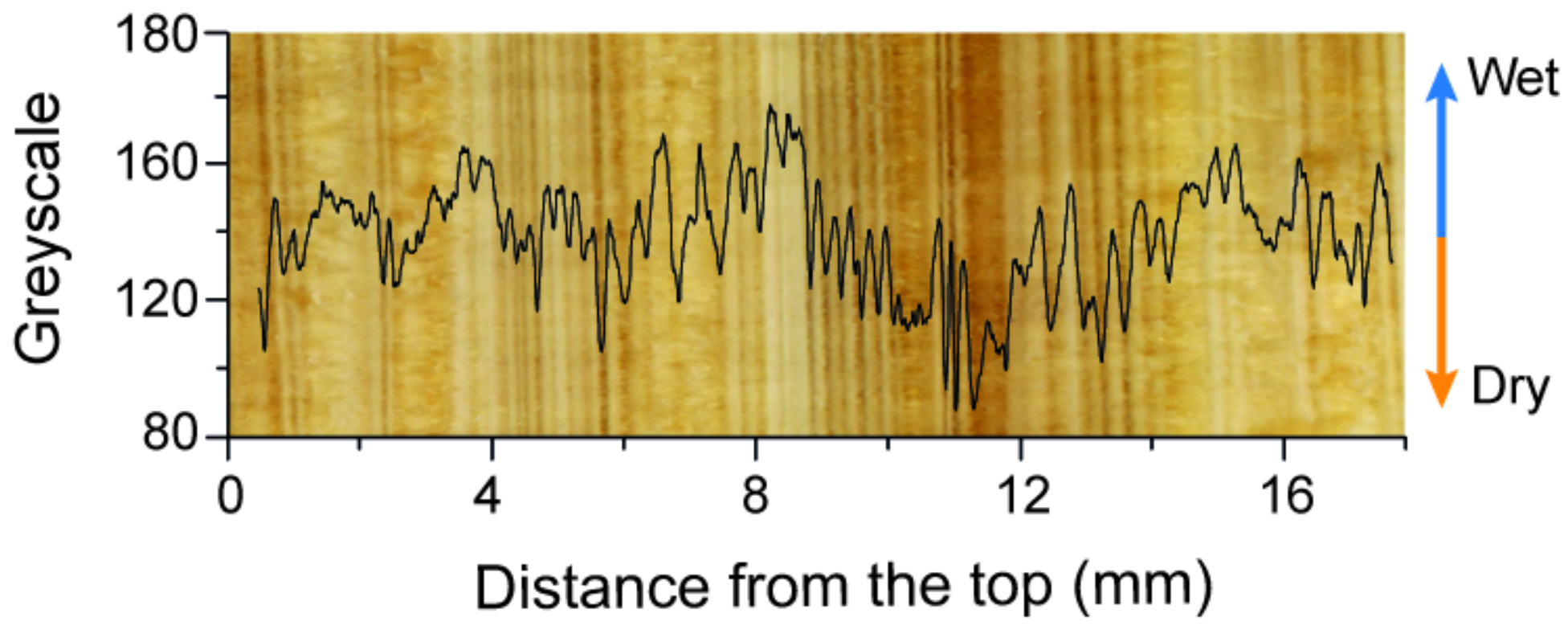}
\caption{Greyscale record (in black) extracted from a high resolution scan of the surface of the stalagmite C132 from Niue island. Lower grey values are associated with dense microcrystalline calcite layers that form during drier periods.}
\label{fig5}
\end{figure}
Finally, we perform the recurrence analysis (Fig.~\ref{fig6}c). In order to characterize seasonal features, the period covered by one segment is fixed as one year. Optimization of deletion/adding costs yields $\Lambda_s = 2$. A window size of $s = 
200\,\mathrm{years}$ is chosen with $90\%$ overlap. A recurrence plot with fixed recurrence rate of $15\%\,$ is obtained for each window 
and analyzed with DET. DET reveals variations in seasonal-scale predictability for the 
real greyscale record (Fig.~\ref{fig6}c, blue line). The effect of the varying sampling size is obtained by
the $95\%\,$-quantile of the DET distribution from 200 SRC-surrogates (Fig.~\ref{fig6}c, yellow line). Five exemplar SRC-surrogate realizations are shown in appendix B.
Both DET time series indicate an increase of seasonal-scale predictability during the wet period between 6.35 to 6.2 ka BP, potentially 
caused by the simultaneously increased sampling rate. The predictability estimate is corrected for the identified sampling bias by considering
the ratio $\mathrm{DET}_\mathrm{r} / \mathrm{DET}_\mathrm{surr}$ (Fig.~\ref{fig6}d).
Two periods (6.4 and 6.2 ka BP, and between 5.9 and 5.72 ka BP) show 
relatively low segment size-corrected seasonal 
predictability $\left. \mathrm{DET}_{\mathrm{r}} \middle/ \mathrm{DET}_{\mathrm{surr}} \right. < 1$.
While the latter is not significantly affected by the correction, the former can only be 
identified as less predictable when the variations in sampling rate are taken into account. This result corroborates 
previous findings that suggested that both of these identified periods were more irregular, i.e. showing less steady seasonal fluctuations \citep{CINTHYA}. 
However, it was not possible to characterize all sub-annual values as a proxy for sub-annual rainfall 
distribution rather extracting only the contrast between wet and dry season. The (m)Edit-distance approach employed here in combination 
with the proposed correction technique allows for a more reliable interpretation of mid-Holocene seasonal variations in the 
west Pacific. 

In particular, an enhanced control of the seasonal cycle by ENSO-scale variability was found in \citep{CINTHYA} for the periods of reduced predictability (6.4 and 6.2 ka BP, and between 5.9 and 5.72 ka BP). High tropical cyclone activity between 6.4-6.2 ka BP could have been triggered by increased ENSO activity, yielding a more irregular sub-annual distribution of rainfall. Our results indicate that not only contrast between both seasons is rendered less predictable during this period but also the seasonal rainfall distribution appears less stable from one year to another. Reconstructing past climate variability at seasonal scale plays a critical role in the context of human adaption to continuous and abrupt climate variations and therefore our approach has direct relevance for teasing out the seasonal-scale signals.
\begin{figure*}[htbp]
\centering
\includegraphics[width=.75\textwidth]{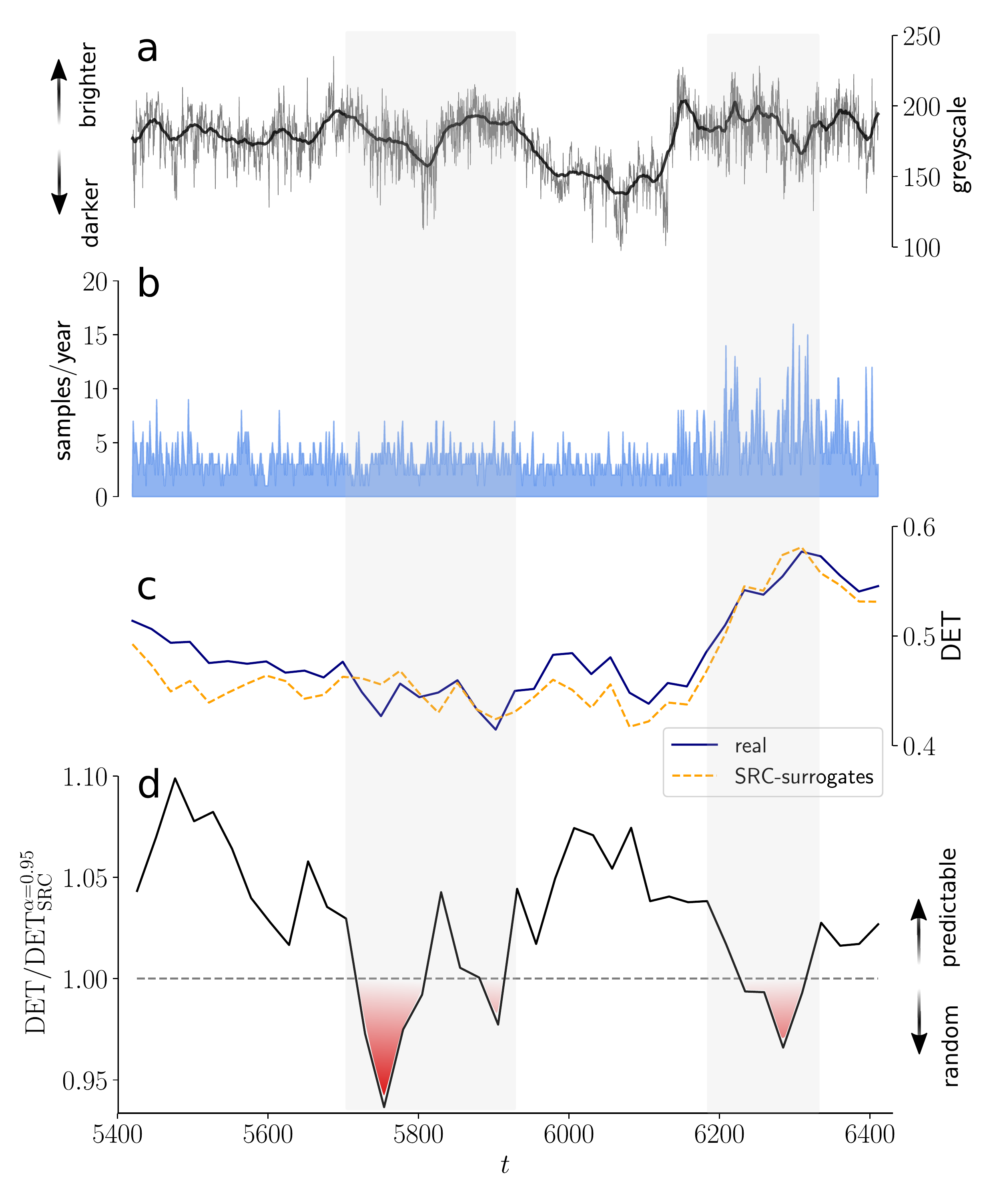}
\caption{Application of SRC-surrogate correction method to an irregularly sampled greyscale proxy-record from the central 
Pacific \textbf{(a)} with non-stationary sampling rate (blue) \textbf{(b)}. A sliding window RQA using determinism (DET) as 
a predictability measure is carried out \textbf{(c)}. Real DET values are displayed in dark blue, the $95\%\,$-confidence 
level computed from 200 SRC-surrogates is shown in yellow. Dividing $\mathrm{DET}_{\mathrm{real}}$ by 
$\mathrm{DET}_{\mathrm{surr}}$ 
yields a sampling-bias corrected predictability series \textbf{(d)} that indicate mid-Holocene variations 
in seasonal-scale predictability. Gray shading indicates two periods with low seasonal predictability.}
\label{fig6}
\end{figure*}
%\FloatBarrier

\section{Conclusion}
\label{sec6}
The characterization of time series from complex nonlinear systems is a challenging task. Irregular sampling, 
i.e., variations in the sampling interval between consecutive values, additionally impedes typical research objectives such 
as spectral analysis, persistence estimation or quantifying the predictability of a system. Even though interpolation 
techniques offer a seemingly efficient way of pre-processing a time series to allow application of standard 
time series analysis tools, these entail various biases which are difficult to control. A different perspective is pursued by 
the (m)Edit-distance method. Many analysis methods are based on an estimate of (dis)similarity. With the 
(m)Edit-distance, 
a suitable distance measure between states of a system at different times $i$ and $j$ is defined by computing the 
transformation cost of segments centered at these time instances. First analyses demonstrated  its scope in the context of 
recurrence analysis, enabling researchers to examine predictability variations of irregularly sampled palaeoclimate time 
series. Applications to other complex systems (also for time series with `missing values') and other methodological 
frameworks (e.g., complex networks, clustering, correlation analysis, \ldots) are possible and should 
be attempted in the near future.

For some real-world systems, it is instructive to quantitatively compare sequences corresponding to a specific time scale in 
order to analyse the scale-specific variability. In such cases, segment sizes will vary in the presence of irregular sampling. 
Furthermore, splitting time series with a non-stationary sampling rate into segments that do not cover the same time period 
will result in a mixing of time scales.
We have shown that (m)Edit-distance-based recurrence analyses are affected by variations in segment sizes, resulting in a 
non-trivial sampling bias if episodes with variable sampling rate are included in a single RP. The (m)Edit distance regards 
pairs of longer segments to be generally more dissimilar than shorter segments due to higher deletion costs. Shifting costs 
conversely decrease for increasing segment size, resulting in a non-trivial dependence of costs on local sampling rates. 
When including amplitudes of a signal into the (m)Edit-distance computation, similar general tendencies 
were observed but the strength of the segment size-dependencies varied for different systems. A more detailed examination of how 
dissimilar amplitude segments of different paradigmatic systems depend on their time scale will be investigated in 
a future study.

We introduced a numerical technique based on constrained randomization to address and correct the issue of segment-size dependence in 
recurrence analysis.
This method involves generating an ensemble of sampling rate constrained surrogate realizations (SRC-surrogates). Each SRC-surrogate 
reproduces the real variations of sampling rate and its assignment to the corresponding amplitude differences, allowing the 
ensemble to be used for correcting the undesired segment size-dependence. The effectiveness of the proposed correction was 
applied to a synthetic AR(1)-time series and real palaeo-proxy data. In both applications, a recurrence analysis 
successfully recovered variations in the scale-specific predictability of the system whilst discarding spurious effects 
imprinted by sampling rate variations. We found that seasonal-scale predictability varied significantly during the mid-Holocene in the 
West Pacific, corroborating and extending the results from a recent study.
The reasons for these changes in predictability warrant further investigation.

The identified sampling bias is a specific case of a more general problem in time series analysis; 
sliding window analyses (or the study of short time series) often suffer from finite-sampling biases which may introduce 
artificial variability into any statistical indicator that is computed. As pointed out in Section \ref{sec5}, finite-sampling biases are also not limited to irregular temporal sampling but are likely to also occur in settings where other parameter axes 
determine suitable window sizes or adaptive windowing is required. In future, the proposed method could also be 
appliedd in such settings to test its effectiveness beyond the examples considered in this study.
\section*{Code availability}
Python code for the generation of SRC-surrogates is available at \url{https://github.com/ToBraun/SRC-surrogates}. 
%%%%%%%%%%%%%%%%%%%%%%%%%%%%%%%%%%%%%%%%%%%%%%%%%%%%%%%%%%%%%%%%%%%%%
%%%%%%%%%%%%%%%%%%%%%%%%%%%%%%%%%%%%%%%%%%%%%%%%%%%%%%%%%%%%%%%%%%%%%
\section*{Acknowledgements}
This research was supported by the
Deutsche Forschungsgemeinschaft in the context of the DFG
project MA4759/11-1 ‘Nonlinear empirical mode analysis of
complex systems: Development of general approach and appli-
cation in climate’. It also received financial support from the European Union’s Horizon 2020 Research and Innovation programme (Marie Sklodowska-Curie grant agreement No. 691037).
Deniz Eroglu acknowledges funding by T\"UB\.ITAK (Grant No.~118C236) and the BAGEP Award of the Science Academy.
Cinthya Nava-Fernandez acknowledges financial support from the German Academic Exchange Service (DAAD). AH also acknowledges support from the Royal Society of New Zealand (grant no. RIS-UOW1501), and the Rutherford Discovery Fellowship programme (grant no. RDF-UOW1601).

\section*{Conflict of interest}
The authors declare that they have no conflict of interest.

\clearpage
\appendix
\section*{Appendix A: analytical and numerical segment size relations}
\label{appA}
In the following, we elaborate on the systematic dependence of the (m)Edit distance on segment lengths $N_a =|\mathcal{S}_{a}|,\, N_b = |\mathcal{S}_{b}|$ in the most simple application: we study events (i.e., no assumptions about the amplitude of the signal) which are unevenly spaced by
exponentially distributed sampling intervals $\Delta$ with a sampling rate $\lambda$:
\begin{align}
p(\Delta,\lambda) \, = \,  \lambda \text{e}^{-\lambda \Delta}
\label{eq6}
\end{align}
Consequently, the number of samples per unit interval $k$ is Poisson-distributed
\begin{align}
\rho(k,\lambda) \, = \,  \frac{\lambda^k\mathrm{e}^{-k}}{k!}
\label{eq7}
\end{align}
with $\lambda$ being equivalent to the expected number of samples per unit interval; $\mathbb{E}(X) = \lambda$. Furthermore, the $n$-th time step is Erlang-distributed with the rate parameter $\lambda$:
\begin{align}
f(t;n,\lambda) \, = \,  \frac{\lambda^n t^{n-1}\mathrm{e}^{-\lambda t}}{(n-1)!}
\label{eq8}
\end{align}
which is a general result for a sum of $n$ independent exponential random variables with 
equivalent rate parameters $\lambda$ \cite{cox1967renewal}.

\begin{figure*}[]
\begin{center}
\hspace{-.5cm}\includegraphics[width=.32\textwidth]{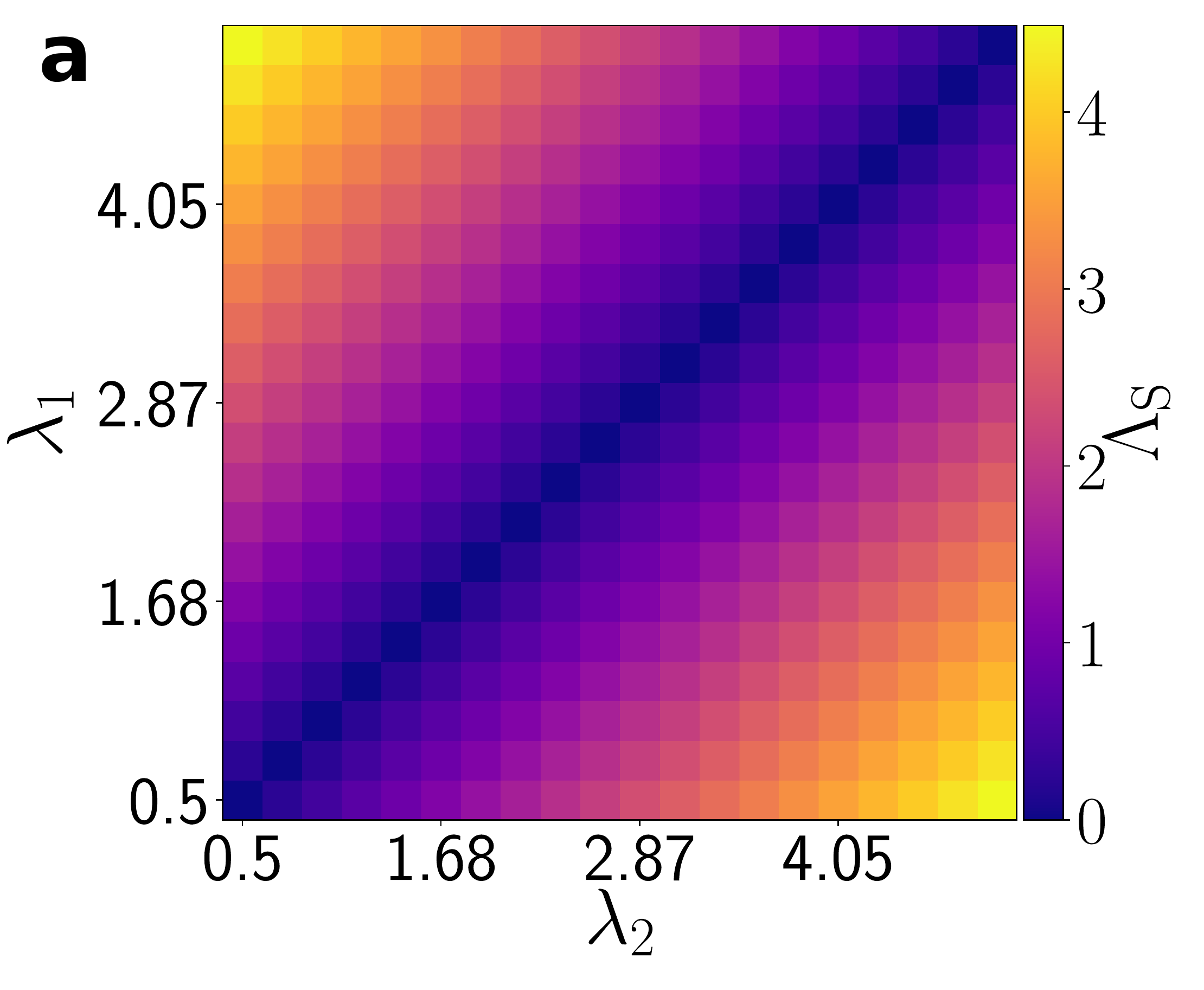}
\includegraphics[width=.32\textwidth]{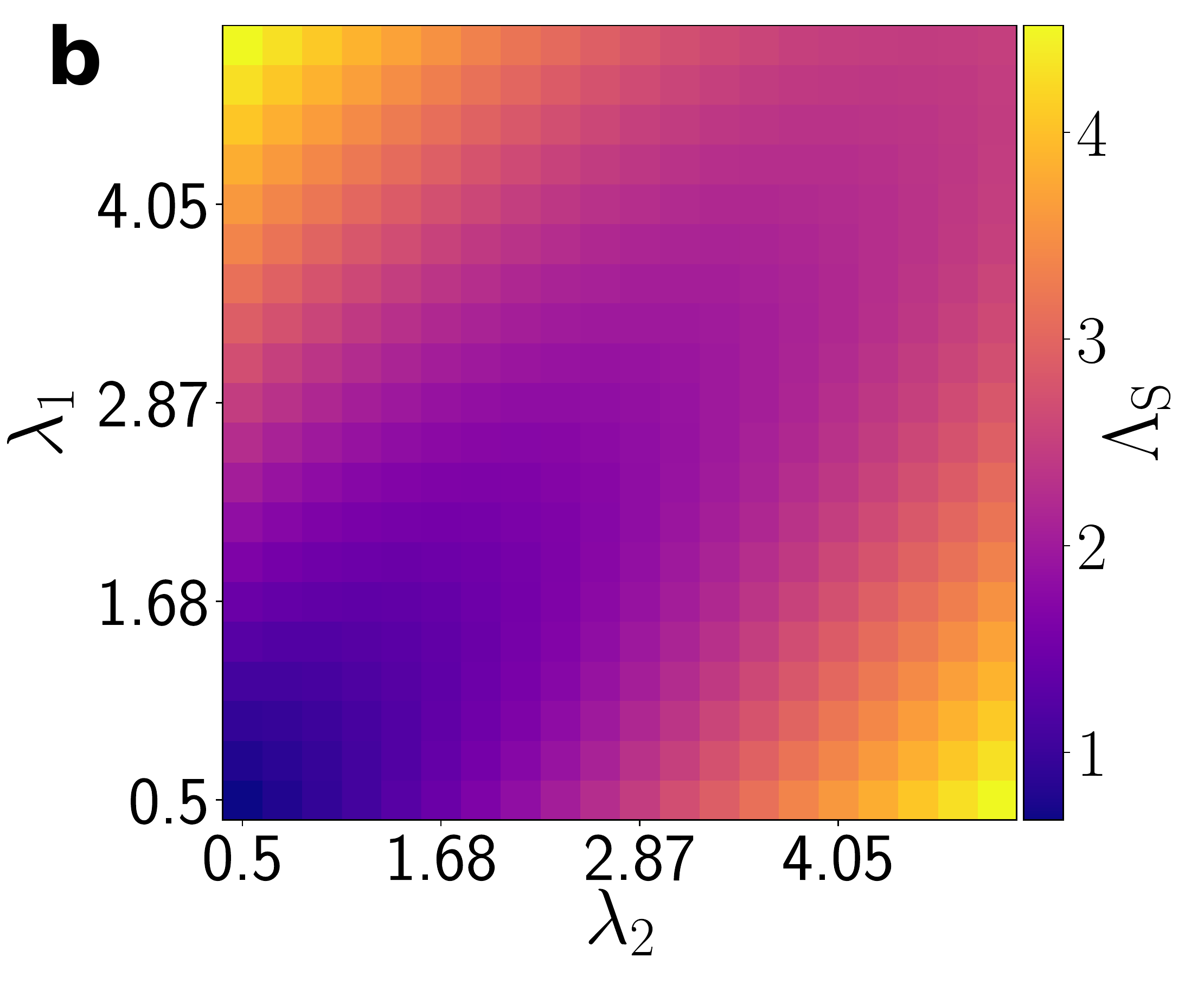}
\includegraphics[width=.26\textwidth]{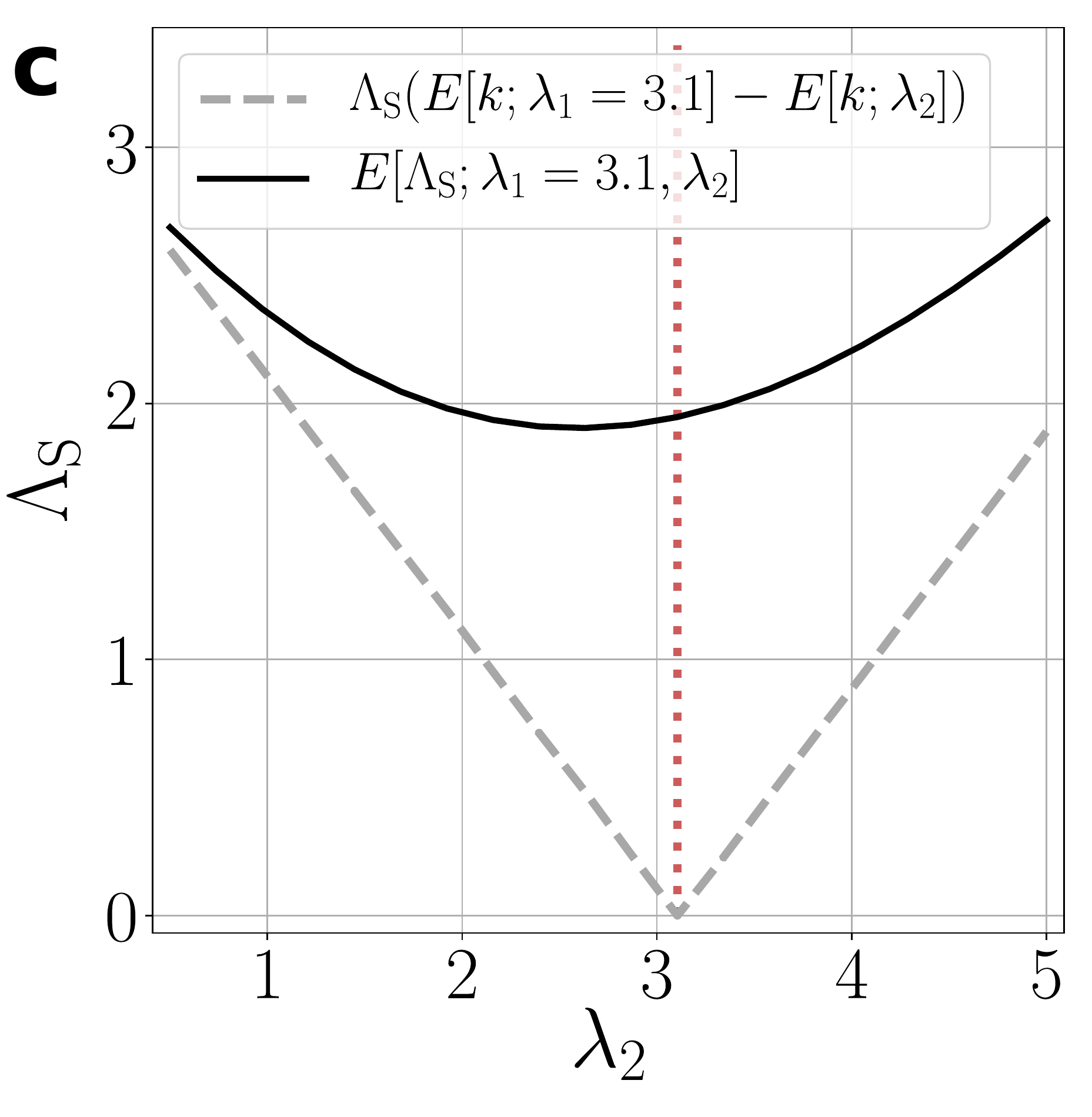}
\includegraphics[width=.32\textwidth]{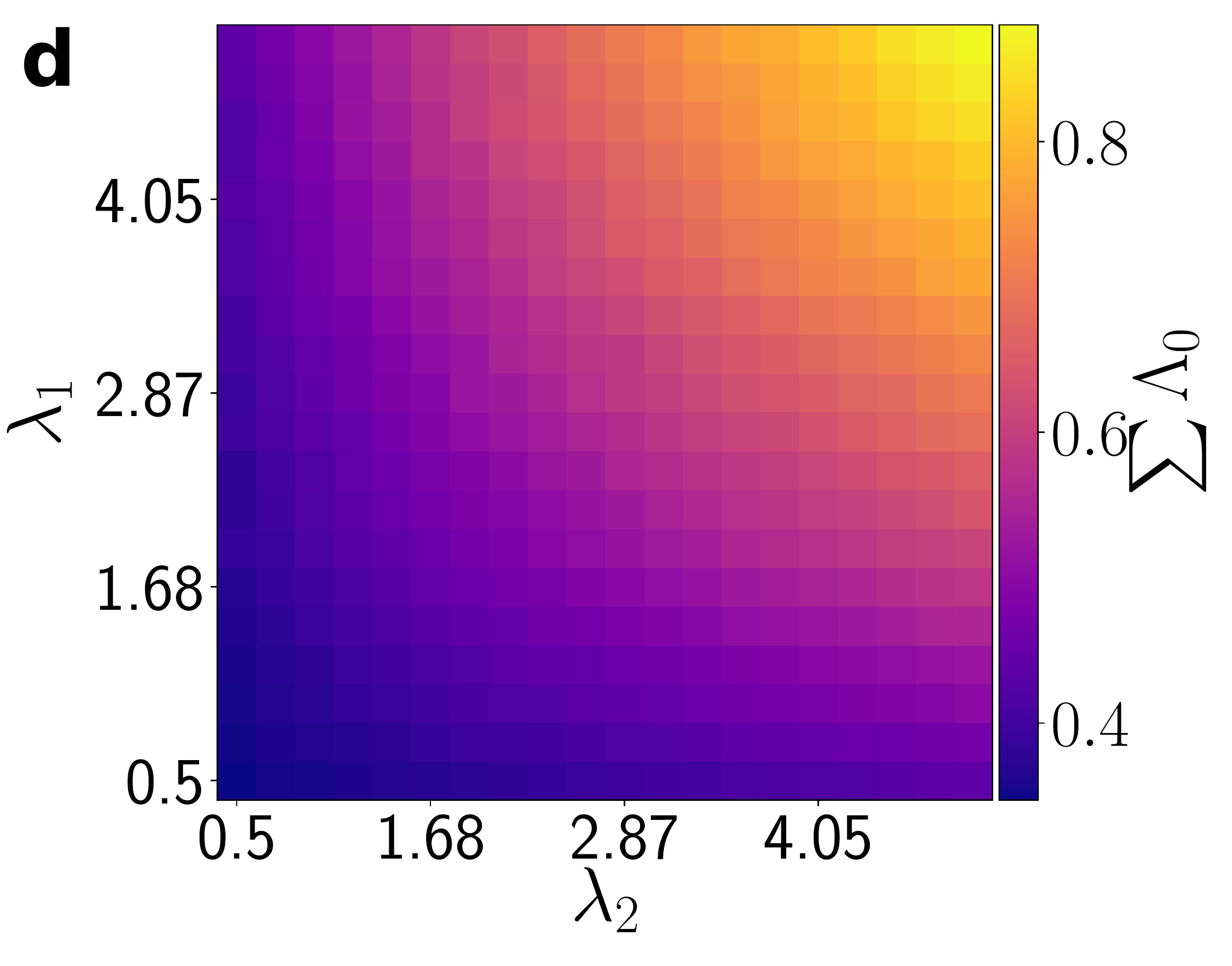}
\includegraphics[width=.32\textwidth]{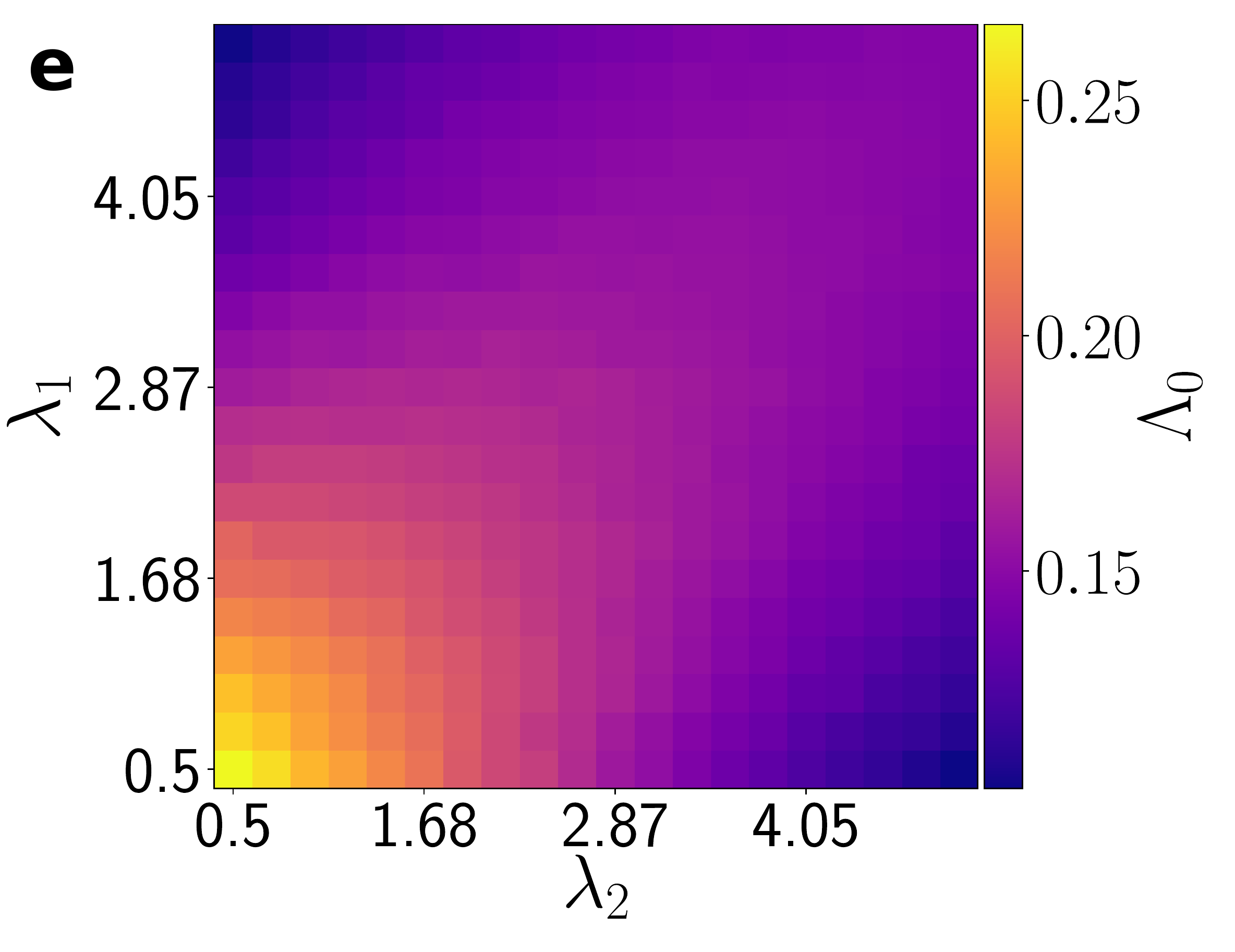}
\includegraphics[width=.28\textwidth]{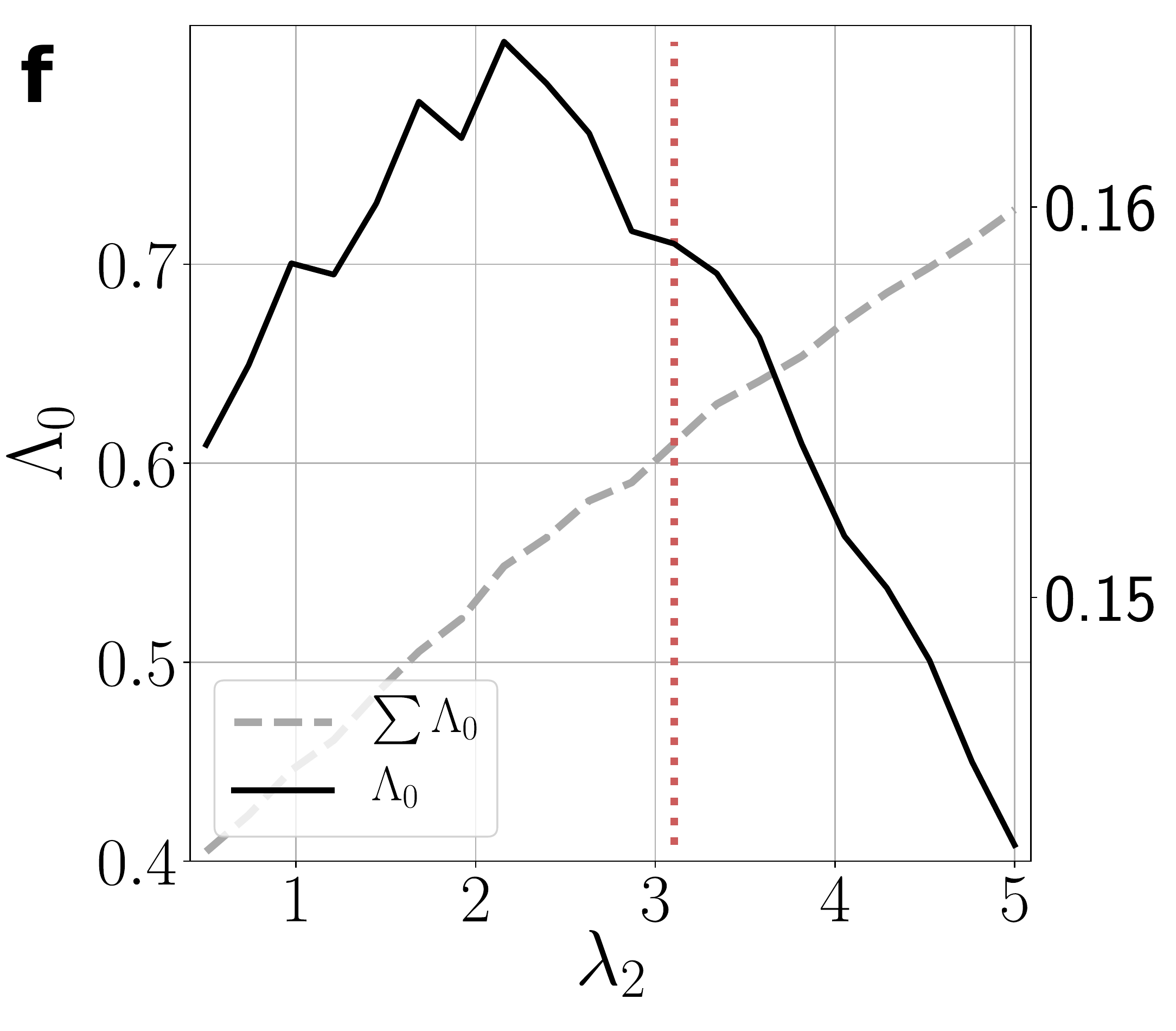}
\includegraphics[width=.32\textwidth]{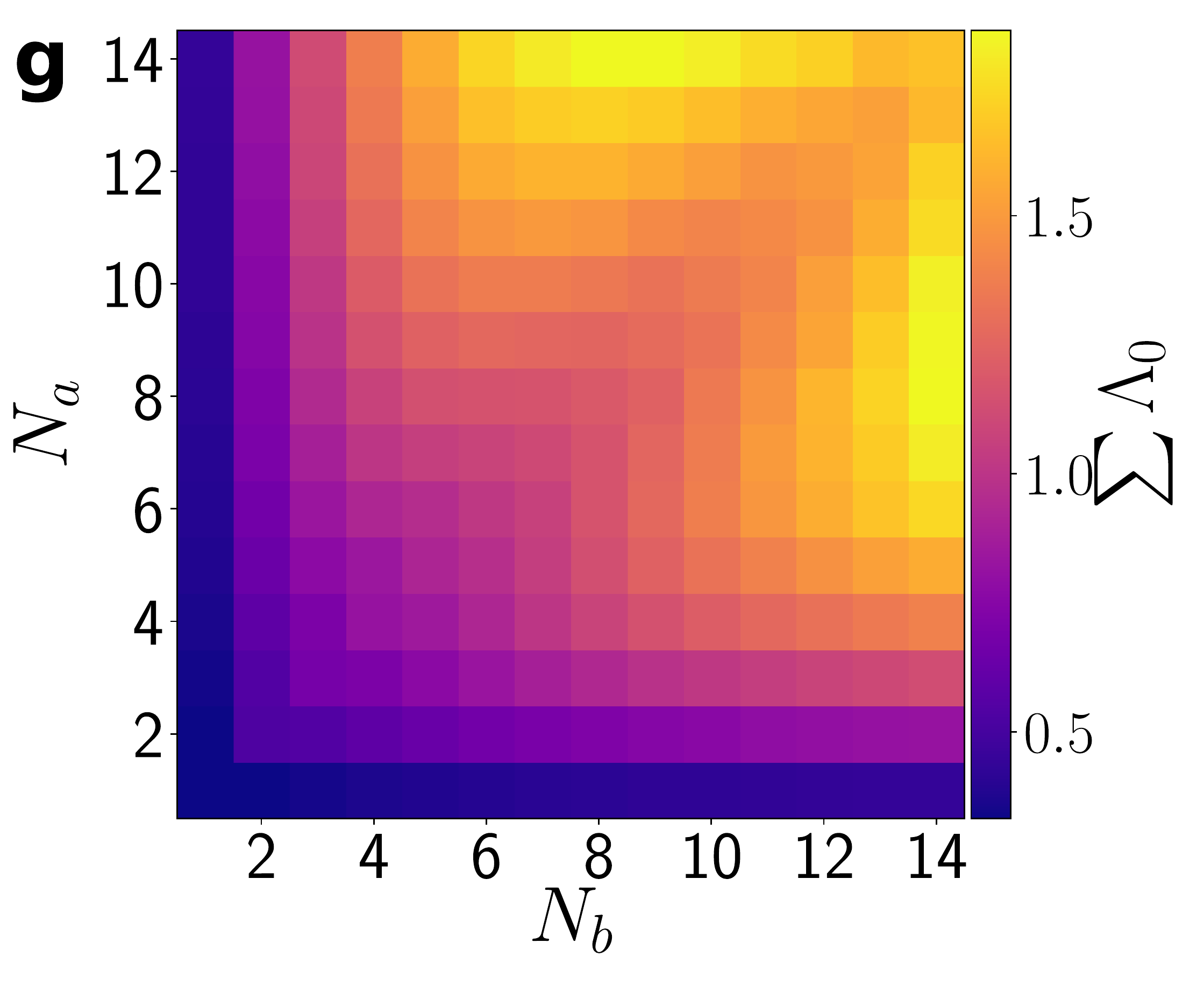}
\includegraphics[width=.32\textwidth]{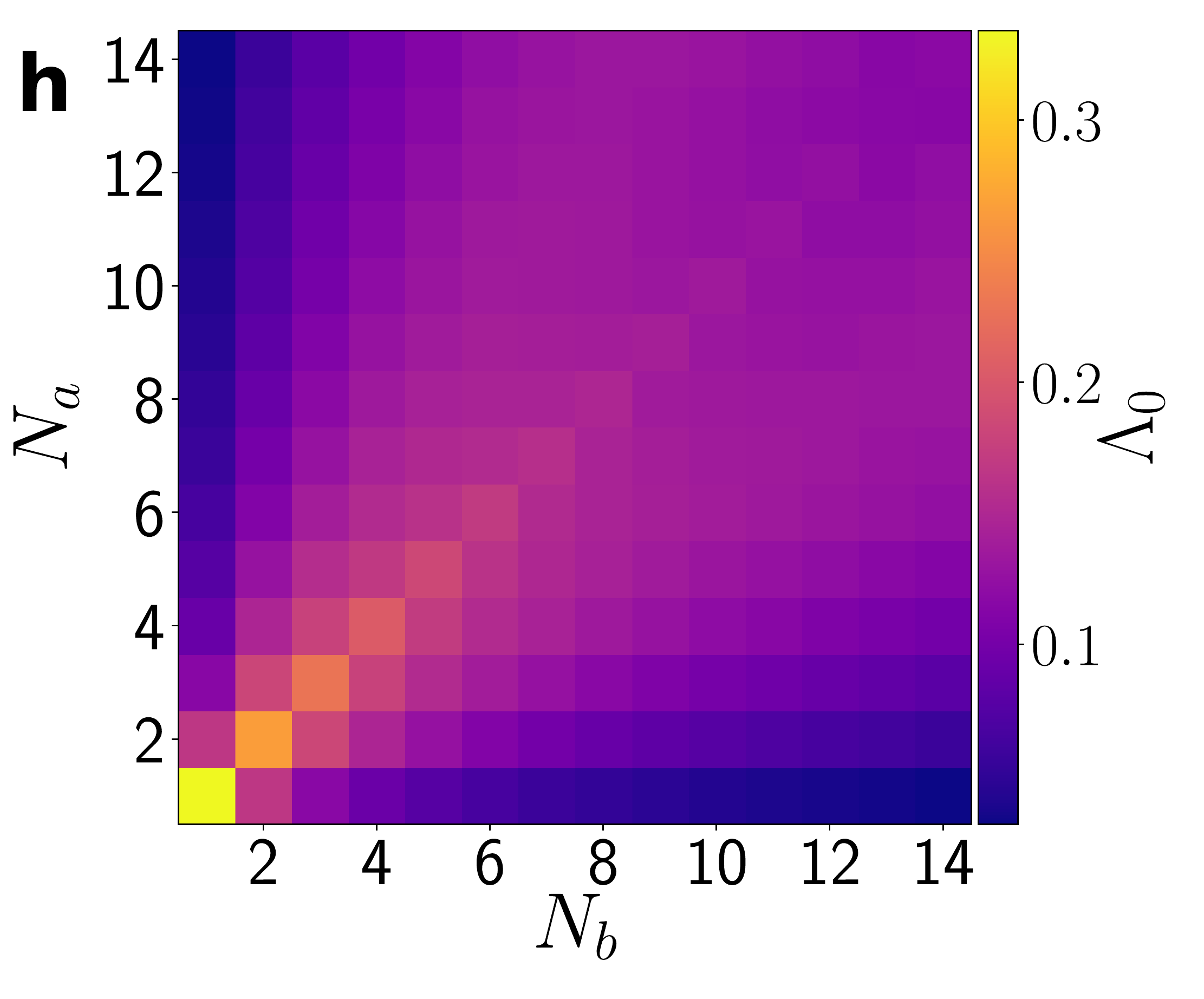}
\includegraphics[width=.28\textwidth]{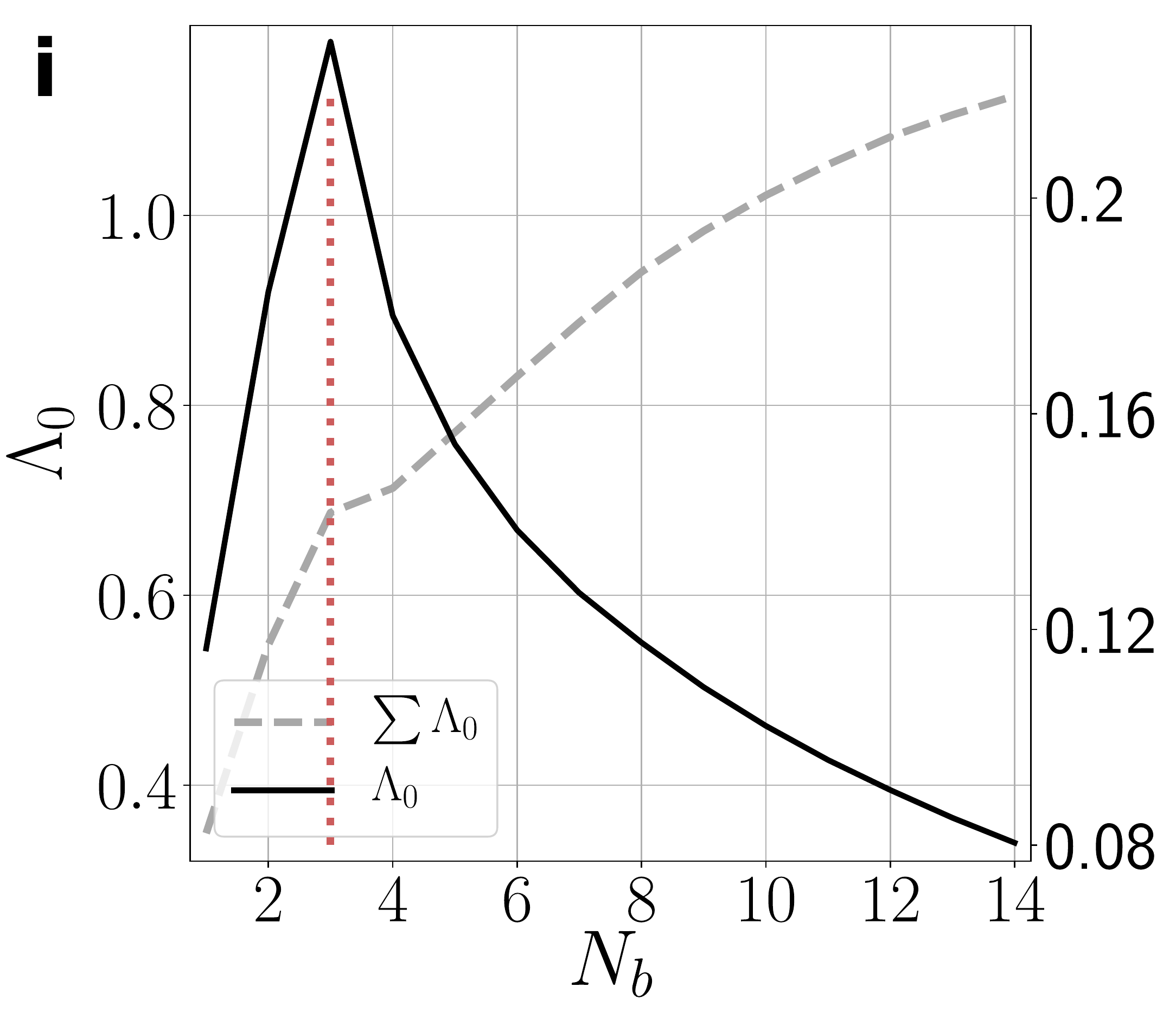}
\end{center}
\caption{
Subcosts for adding/deleting (a-c) and shifting (d-i) operations for exponentially distributed 
sampling intervals. Only necessary deleting/adding operations are regarded (`no competing 
operations').
The sampling rate-dependence of deletion costs is given as difference between 
expected number of samples per unit interval (left matrix and grey dashed line) and as 
expected costs given two rates $\lambda_1, \lambda_2$ (right matrix and black line, eq.~(\ref{eq15})).
Shifting costs are 
studied (b) numerically with respect to their dependence on the sampling rates $\lambda_ 1,\, 
\lambda_2$ and  (c) on the actual number of samples per unit interval $N_a,\,N_b$.
The left matrices shows shifting costs for the full transformation of segments, the center matrices show shifting costs per operation.
Exemplary columns are displayed in the line plots whereas the red dashed line marks the respective rate $\lambda_1$/segment size $N_b$.}
\label{figA1}
\end{figure*}

We are interested in the segment size-dependence of deletion(/adding) and shifting costs for the 
edit distance. This can be evaluated by considering $M$ exponential random variables where 
each is drawn from a distribution $p(\Delta,\lambda_m)$ with distinct 
$\lambda_m,\,m=1,2,\dots,M$. 
When applied, this setting can be considered equivalent to a 
scenario where a time axis changes its local sampling rate $\lambda_m$ at $M$ points and 
segments from these should be compared via the edit-distance. For a specific pair of segments 
with sizes $N_a, N_b$, the minimum deletion cost (no deletions as competing to shifts 
included) for their transformation is
\begin{align}
C_{\mathrm{del}}(N_a, N_b) \, = \, \Lambda_{\mathrm{S}}|N_a - N_b|
\label{eq9}
\end{align}
Consequently, for two segments of average sizes $\mathbb{E}[N_a] = \lambda_1,\,\mathbb{E}[N_b] 
= \lambda_2$ we obtain a minimum deletion cost of $C_{\mathrm{del}}\left(\mathbb{E}(N_a), 
\mathbb{E}(N_b)\right) 
\, = \, \Lambda_{\mathrm{S}}|\lambda_1 - \lambda_2|$. A cost matrix $C_{\mathrm{del}}(\lambda_1,\lambda_2)$ is exemplified in Fig.~\ref{figA1}a.
The expected minimum deletion cost for two randomly selected segments from time periods with 
different rates $\lambda_1,\,\lambda_2$ can be computed by using the the Skellam distribution
\begin{widetext}
\begin{align}
\rho_s(k=|z|;\lambda_1,\lambda_2) \, = \, 
\left\{
\begin{array}{ll} 
\mathrm{e}^{-\lambda_1-\lambda_2}\left((\frac{\lambda_1}{\lambda_2})^{\frac{k}{2}}I_k(2\sqrt{\lambda_1\lambda_2}) \,+\,  (\frac{\lambda_2}{\lambda_1})^{\frac{k}{2}}I_{-k}(2\sqrt{\lambda_1\lambda_2}) \right)
& \mathrm{if}\, k>0 \\
\mathrm{e}^{-\lambda_1-\lambda_2} I_0(2\sqrt{\lambda_1\lambda_2})
& \mathrm{if}\, k=0
\end{array}
\right.
\label{eq10}
\end{align}
\end{widetext}
for the difference $Z = X-Y$ where $X,\,Y$ are Poisson-distributed random variables with rates 
$\lambda_1,\,\lambda_2$, Eq.~(\ref{eq7}). $I_k(a)$ denotes the modified Bessel function of the first kind. 
For $k>0$, the moment-generating function is consequently given by
\begin{align}
\begin{split}
M(t; \lambda_1, \lambda_2) \, = \, &\mathrm{e}^{-\lambda_1-\lambda_2}
\left( \sum_{k=0}^{\infty} \mathrm{e}^{tk}I_k(2\sqrt{\lambda_1\lambda_2})
\right.
\\
& \left. \left[
\left(\frac{\lambda_1}{\lambda_2}^{k/2} \right)
+ \left(\frac{\lambda_2}{\lambda_2}^{k/2} \right)
\right]
- I_0(2\sqrt{\lambda_1\lambda_2})
\right)
\end{split}
\label{eq11}
\end{align}
With `Marcum's Q'
\begin{align}
Q(\sqrt{2\lambda_1},\sqrt{2\lambda_2}) = \mathrm{e}^{-\lambda_1-\lambda_2}\sum_{k=0}^{\infty} \left(\frac{\lambda_1}{\lambda_2}\right)^{\frac{k}{2}}I_k(2\sqrt{\lambda_1\lambda_2})
\label{eq12}
\end{align}
and its derivative
\begin{align}
\begin{split}
\dfrac{\mathrm{d}}{\mathrm{d}t}Q(\sqrt{2\lambda_1},\sqrt{2\lambda_2})
\, = \, \mathrm{e}^{-\lambda_1\mathrm{e}^t -\lambda_2\mathrm{e}^{-t}}
\left(
\lambda_2\mathrm{e}^{-t} I_0(2\sqrt{\lambda_1\lambda_2}) + \right. \\
\left. \sqrt{\lambda_1\lambda_2}I_1(2\sqrt{\lambda_1\lambda_2})
\right)
\end{split}
\label{eq13}
\end{align}
this can be written as

\begin{align}
\begin{split}
M(t; \lambda_1, \lambda_2) \, = \, \mathrm{e}^{-\lambda_1-\lambda_2}
\left[
Q\left(\sqrt{2\lambda_2\mathrm{e}^{-t}}, \sqrt{2\lambda_1\mathrm{e}^{t}}\right)
\mathrm{e}^{\lambda_1 \mathrm{e}^t
+ \lambda_2 \mathrm{e}^{-t}}
\right. \\
\left. Q
\left
(\sqrt{2\lambda_1\mathrm{e}^{-t}}, \sqrt{2\lambda_2\mathrm{e}^{t}}
\right)
\mathrm{e}^{\lambda_2 \mathrm{e}^t + \lambda_1 \mathrm{e}^{-t}}
I_0(2\sqrt{\lambda_1\lambda_2})
\right]
\end{split}
\label{eq14}
\end{align}
Differentiating this moment-generating function (using eq. \ref{eq13}) around $t=0$ with Leibniz rule yields the expected value:

\begin{align}
\begin{split}
\mathbb{E}[k; \lambda_1, \lambda_2] \ = \ &2\mathrm{e}^{-\lambda_1-\lambda_2}
\left(
\lambda_2I_0(2\sqrt{\lambda_1\lambda_2})
+ \sqrt{\lambda_1\lambda_2}I_1(2\sqrt{\lambda_1\lambda_2})
\right) 
\\
\, + \, 
&\left(
\lambda_2 - \lambda_1)(1-2Q(\sqrt{2\lambda_1},\sqrt{2\lambda_2})
\right)
\end{split}
\label{eq15}
\end{align}
Hence, $\mathbb{E}[C_{\mathrm{del}}(\lambda_1, \lambda_2)] \, = \, \Lambda_{\mathrm{S}}\mathbb{E}[k; \lambda_1, \lambda_2]$ (Fig.~\ref{figA1}a (middle)).
In the right line plot of Fig.~\ref{figA1}a, two columns with $\lambda_1$ fixed at $3.1$ are shown to 
illustrate the scaling of deletion costs with the rate $\lambda$ more clearly. While 
$C_{\mathrm{del}}\left(\mathbb{E}(N_a), \mathbb{E}(N_b)\right)$ shows a sharp minimum at the 
rate $\lambda_2=\lambda_1$, $\mathbb{E}[k; \lambda_1, \lambda_2]$ decreases more smoothly with 
increasing $\lambda_2$, and increases afterwards. The latter becomes minimal for a value 
$\lambda<\lambda_2$ 
instead of $\lambda=\lambda_2$ since Poisson-distributions $\rho(k,\lambda)$ are right-skewed, having higher cumulated probability mass for all values $k>\lambda$.
Note that all said above holds in the same way for adding operations.

For the analysis of shifting costs, we focus on the simple case of linear shifting costs
\begin{align}
\tilde{f}_{\Lambda_0}\left(t(\alpha), t(\beta)\right) \, = \, |t(\alpha) - t(\beta)|
\label{eq16}
\end{align}
between the $\alpha$-th event in segment $\mathcal{S}_a$ and the $\beta$-th event in segment 
$\mathcal{S}_b$ 
as proposed in the original, unmodified edit-distance measure. To exclude effects caused by absolute timing of events, timing of events within each segment is always 
transformed into the interval $I=[0,1]$.
The sum of all shifting costs for a pair of segments is denoted as $d_{ab} = 
\Lambda_0\sum_{\alpha,\beta} f_{\Lambda_0}\left(t(\alpha), t(\beta)\right)$ with 
$\Lambda_0=1$. Note that $N_a=N_b$ as $|N_a-N_b|$ deletions/addings have 
already been carried out.
A closed-form solution for the shifting costs between two time instances drawn randomly from 
the distributions $f(x;m_1,\lambda_1),\, f(y;m_2,\lambda_2)$ most likely exists, at least for 
the case $m_1=m_2$ but its computation is beyond this study. We examine shifting costs for 
this case numerically, while we explicitly exclude any deletions as an 
alternative operation to shifting after the neccessary $|N_a-N_b|$ deletions (`basic 
deletions') (Fig.~\ref{figA1}b). We fix $w=1$ as the unit interval 
(arbitrary units). The numerical estimate of the average cost for transforming a segment 
sampled with rate $\lambda_1$ into a segment sampled with rate $\lambda_2$ is based on 
generating time axes for a fixed time period $T=10,000$ (but varying number of events). Given 
a fixed combination of $\lambda_1,\,\lambda_2$, a total of $K=10,000$ segment pairs are 
randomly sampled (with replacement) from both corresponding time axes. The edit-distance is 
computed for each pair of segments and averaged over all pairs to obtain a single value 
$\overline{d}(\lambda_1,\,\lambda_2)$ 
that is characteristic for the combination of rates 
$\lambda_1,\,\lambda_2$. 
This is shown as a cost matrix 
$C_{\mathrm{shift}}(\lambda_1,\lambda_2)$ 
of averaged total shifting costs between randomly drawn segments (Fig.~\ref{figA1}b, left).
The total number of shifts performed after deleting $|N_a - N_b|$ events generally differs for distinct pairs of segments $\mathcal{S}_{a}, \mathcal{S}_{b}$ at fixed 
$\lambda_1,\, 
\lambda_2$. However, when averaged over all randomly drawn segment pairs, an increasing trend along the diagonals is observed. Furthermore, average total shifting 
costs $\overline{d}(\lambda_1,\,\lambda_2)$ increase for fixed $\lambda_1$ and increasing 
$\lambda_2$ (Fig.\ref{figA1}b, right) which is to be expected as a higher 
number of shifts will entail higher summed costs. On the other hand, no monotonous relation between the average shifting costs per shifting operation
\begin{align}
\left.{{\tilde{C}}}_{\mathrm{shift}}(\lambda_1,\lambda_2) = \sum_{k=1}^K d(\mathcal{S}_{a,k}^{(\lambda_1)},\,\mathcal{S}_{b,k}^{(\lambda_2)}) \middle/ \mathrm{max}\{N_a,N_b\} \right. 
\label{eq17}
\end{align}
and sampling rate is observed (Fig.~\ref{figA1}b, center and bold black line on the right). With increasing 
sampling rates, the cost of an average single shifting operation decreases (diagonals of the matrix). For fixed $\lambda_1$, it is maximized at a value $\lambda_2<\lambda_1$ for the same reason as above, i.e. the Erlang distribution being right-skewed. 

If we instead examine the dependence of shifting costs on the actual segment size 
${{\tilde{C}}}_{\mathrm{shift}}(N_a,\,N_b)$ 
rather than the rates (Fig.~\ref{figA1}c), a sharp maximum at $N_b = \lambda_1$ is found (black line, right plot). 
Total shifting costs increase for $N_b < \lambda_1$ and continue to increase more slowly for 
$N_b > \lambda_1$. For fixed $N_a$, an increasing number $N_b$ of events per unit interval 
increases the likelihood that some events are placed close to the events in segment $\mathcal{S}_a$, resulting in lower distances $d_{ab}(N_a,\,N_b)$.
\FloatBarrier

\newpage
\section*{Appendix B: sampling rate-constrained surrogates}
\label{appB}
\begin{figure*}[ht]
\begin{center}
\includegraphics[width=.48\textwidth]{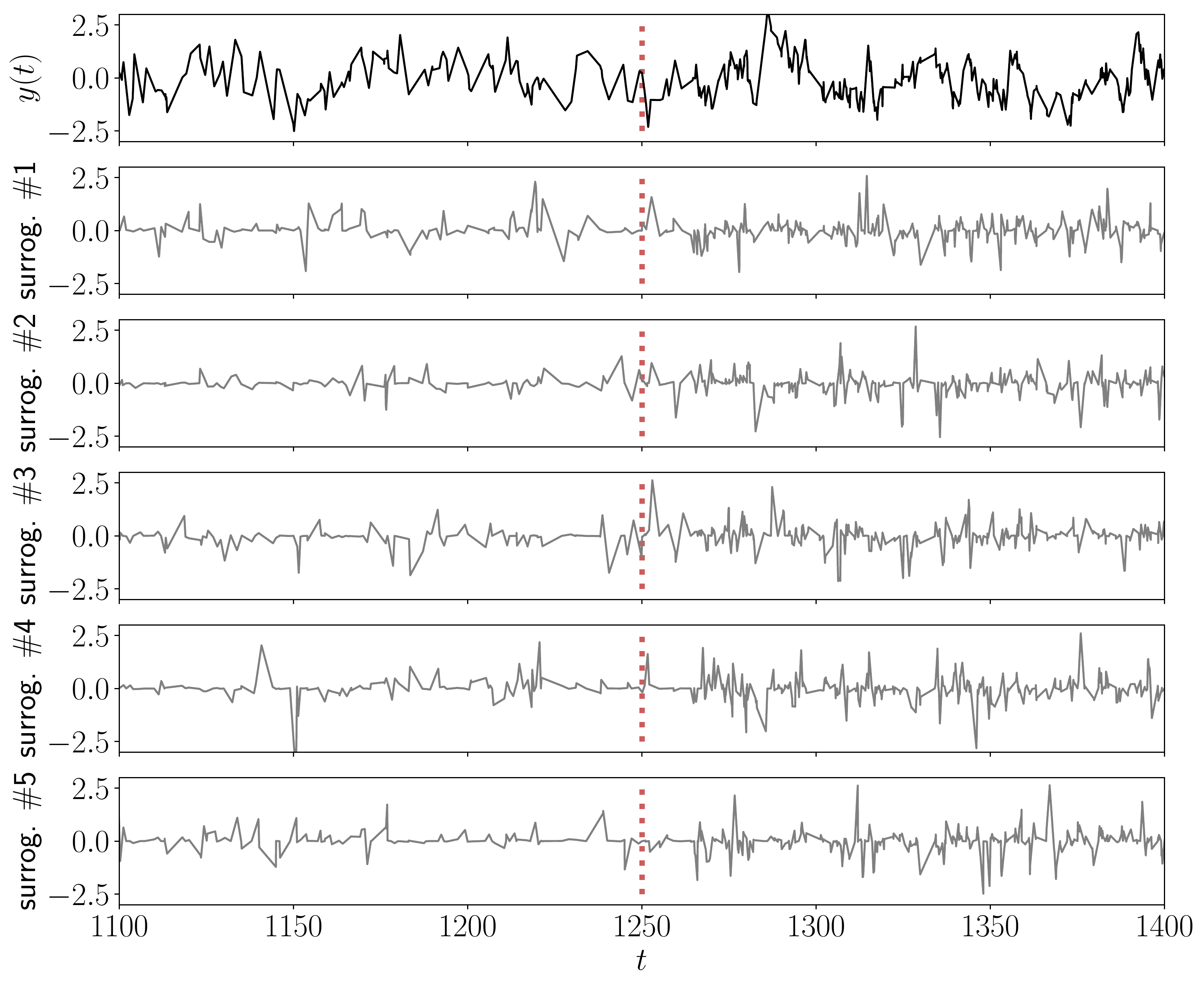}
\caption{Zoomed section of synthetic AR(1)-time series (black) and five exemplar SRC-surrogate realizations (gray). The red dotted line indicates the transition of the sampling rate towards more dense sampling. Sampling intervals are $\gamma\,$-distributed.}
\label{figA2}
\end{center}
\end{figure*}
\begin{figure*}[ht]
\begin{center}
\includegraphics[width=.48\textwidth]{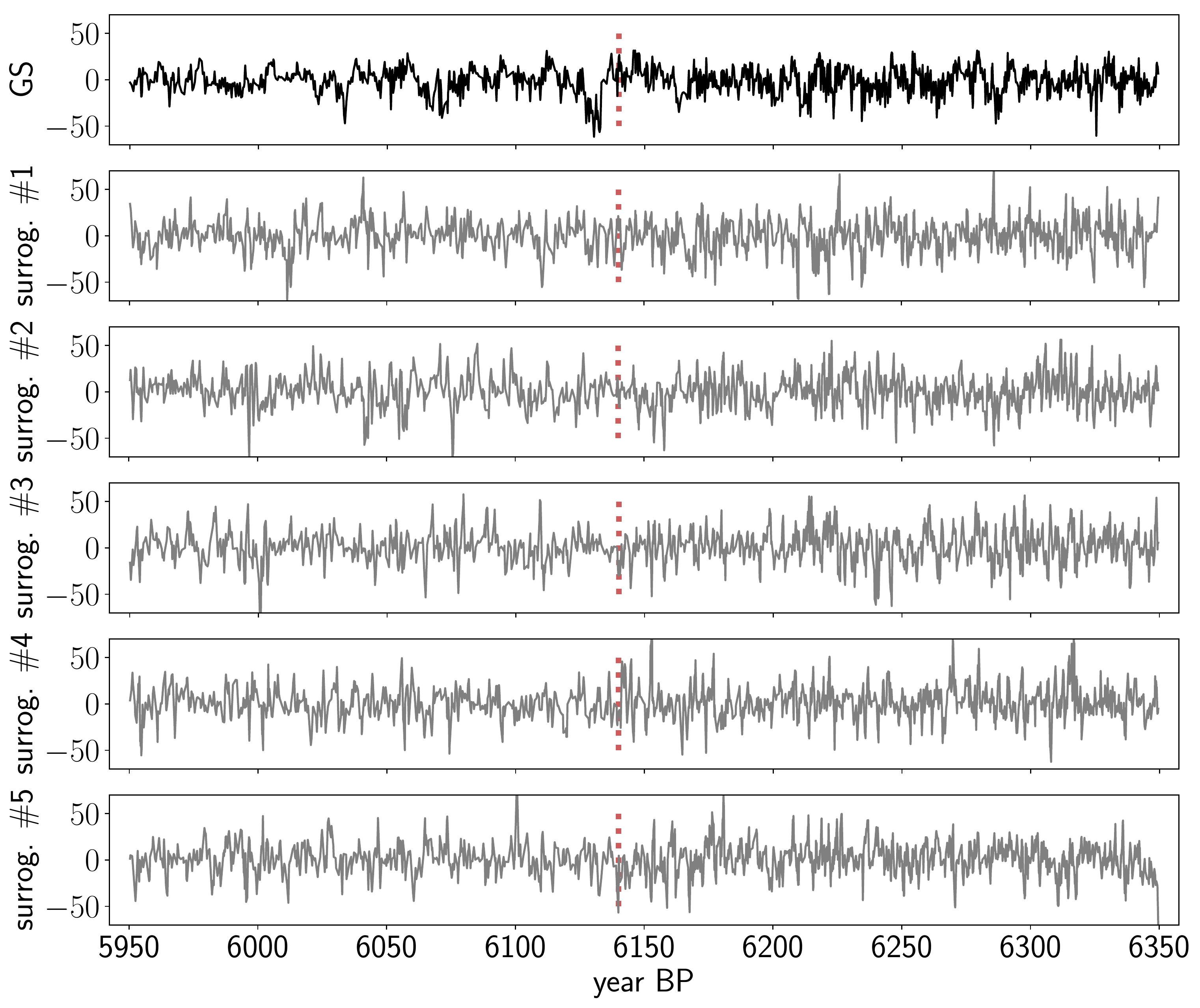}
\caption{Zoomed section of grayscale anomaly time series (black) and five exemplar SRC-surrogate realizations (gray). The red dotted line indicates the transition of the sampling rate towards more dense sampling.}
\label{figA3}
\end{center}
\end{figure*}
\FloatBarrier

\bibliographystyle{ieeetr}
\bibliography{bibli}  %%% Uncomment this line and comment out the 

\begin{thebibliography}{10}

\bibitem{reno2003closer}
R.~Ren{\`o}, ``A closer look at the epps effect,'' {\em International Journal
  of theoretical and applied finance}, vol.~6, no.~01, pp.~87--102, 2003.

\bibitem{fedotov2020methods}
A.~Fedotov, S.~Akulov, and E.~Timchenko, ``Methods of mathematical analysis of
  heart rate variability,'' {\em Biomedical Engineering}, vol.~54, no.~3,
  pp.~220--225, 2020.

\bibitem{enders2010applied}
C.~K. Enders, {\em Applied missing data analysis}.
\newblock Guilford press, 2010.

\bibitem{khayati2020mind}
M.~Khayati, A.~Lerner, Z.~Tymchenko, and P.~Cudr{\'e}-Mauroux, ``Mind the gap:
  an experimental evaluation of imputation of missing values techniques in time
  series,'' {\em Proceedings of the VLDB Endowment}, vol.~13, no.~5,
  pp.~768--782, 2020.

\bibitem{scargle1982studies}
J.~D. Scargle, ``Studies in astronomical time series analysis. ii-statistical
  aspects of spectral analysis of unevenly spaced data,'' {\em The
  Astrophysical Journal}, vol.~263, pp.~835--853, 1982.

\bibitem{rehfeld2011comparison}
K.~Rehfeld, N.~Marwan, J.~Heitzig, and J.~Kurths, ``Comparison of correlation
  analysis techniques for irregularly sampled time series,'' {\em Nonlinear
  Processes in Geophysics}, vol.~18, no.~3, pp.~389--404, 2011.

\bibitem{baldini2021detecting}
J.~U. Baldini, ``Detecting and quantifying paleoseasonality in stalagmites
  using geochemical and modelling approaches,'' {\em AGUFM}, vol.~2017,
  pp.~PP54B--09, 2021.

\bibitem{muhlinghaus2007modelling}
C.~M{\"u}hlinghaus, D.~Scholz, and A.~Mangini, ``Modelling stalagmite growth
  and $\delta$13c as a function of drip interval and temperature,'' {\em
  Geochimica et Cosmochimica Acta}, vol.~71, no.~11, pp.~2780--2790, 2007.

\bibitem{garland2018anomaly}
J.~Garland, T.~R. Jones, M.~Neuder, V.~Morris, J.~W. White, and E.~Bradley,
  ``Anomaly detection in paleoclimate records using permutation entropy,'' {\em
  Entropy}, vol.~20, no.~12, p.~931, 2018.

\bibitem{trauth2021spectral}
M.~H. Trauth, ``Spectral analysis in quaternary sciences,'' {\em Quaternary
  Science Reviews}, vol.~270, p.~107157, 2021.

\bibitem{rehfeld2013late}
K.~Rehfeld, N.~Marwan, S.~F. Breitenbach, and J.~Kurths, ``Late holocene asian
  summer monsoon dynamics from small but complex networks of paleoclimate
  data,'' {\em Climate dynamics}, vol.~41, no.~1, pp.~3--19, 2013.

\bibitem{schulz1997spectrum}
M.~Schulz and K.~Stattegger, ``Spectrum: Spectral analysis of unevenly spaced
  paleoclimatic time series,'' {\em Computers \& Geosciences}, vol.~23, no.~9,
  pp.~929--945, 1997.

\bibitem{chawla2002smote}
N.~V. Chawla, K.~W. Bowyer, L.~O. Hall, and W.~P. Kegelmeyer, ``Smote:
  synthetic minority over-sampling technique,'' {\em Journal of artificial
  intelligence research}, vol.~16, pp.~321--357, 2002.

\bibitem{barua2012mwmote}
S.~Barua, M.~M. Islam, X.~Yao, and K.~Murase, ``Mwmote--majority weighted
  minority oversampling technique for imbalanced data set learning,'' {\em IEEE
  Transactions on knowledge and data engineering}, vol.~26, no.~2,
  pp.~405--425, 2012.

\bibitem{lenton2019climate}
T.~M. Lenton, J.~Rockstr{\"o}m, O.~Gaffney, S.~Rahmstorf, K.~Richardson,
  W.~Steffen, and H.~J. Schellnhuber, ``Climate tipping points—too risky to
  bet against,'' 2019.

\bibitem{bradley2015nonlinear}
E.~Bradley and H.~Kantz, ``Nonlinear time-series analysis revisited,'' {\em
  Chaos: An Interdisciplinary Journal of Nonlinear Science}, vol.~25, no.~9,
  p.~097610, 2015.

\bibitem{marwan2021nonlinear}
N.~Marwan, J.~F. Donges, R.~V. Donner, and D.~Eroglu, ``Nonlinear time series
  analysis of palaeoclimate proxy records,'' {\em Quaternary Science Reviews},
  vol.~274, p.~107245, 2021.

\bibitem{scheffer2009early}
M.~Scheffer, J.~Bascompte, W.~A. Brock, V.~Brovkin, S.~R. Carpenter, V.~Dakos,
  H.~Held, E.~H. Van~Nes, M.~Rietkerk, and G.~Sugihara, ``Early-warning signals
  for critical transitions,'' {\em Nature}, vol.~461, no.~7260, pp.~53--59,
  2009.

\bibitem{lekscha2018phase}
J.~Lekscha and R.~V. Donner, ``Phase space reconstruction for non-uniformly
  sampled noisy time series,'' {\em Chaos: An Interdisciplinary Journal of
  Nonlinear Science}, vol.~28, no.~8, p.~085702, 2018.

\bibitem{mccullough2016counting}
M.~McCullough, K.~Sakellariou, T.~Stemler, and M.~Small, ``Counting forbidden
  patterns in irregularly sampled time series. i. the effects of
  under-sampling, random depletion, and timing jitter,'' {\em Chaos: An
  Interdisciplinary Journal of Nonlinear Science}, vol.~26, no.~12, p.~123103,
  2016.

\bibitem{sakellariou2016counting}
K.~Sakellariou, M.~McCullough, T.~Stemler, and M.~Small, ``Counting forbidden
  patterns in irregularly sampled time series. ii. reliability in the presence
  of highly irregular sampling,'' {\em Chaos: An Interdisciplinary Journal of
  Nonlinear Science}, vol.~26, no.~12, p.~123104, 2016.

\bibitem{suzuki2010definition}
S.~Suzuki, Y.~Hirata, and K.~Aihara, ``Definition of distance for marked point
  process data and its application to recurrence plot-based analysis of
  exchange tick data of foreign currencies,'' {\em International Journal of
  Bifurcation and Chaos}, vol.~20, no.~11, pp.~3699--3708, 2010.

\bibitem{ozken2015transformation}
I.~Ozken, D.~Eroglu, T.~Stemler, N.~Marwan, G.~B. Bagci, and J.~Kurths,
  ``Transformation-cost time-series method for analyzing irregularly sampled
  data,'' {\em Physical Review E}, vol.~91, no.~6, p.~062911, 2015.

\bibitem{ukkonen1985algorithms}
E.~Ukkonen, ``Algorithms for approximate string matching,'' {\em Information
  and control}, vol.~64, no.~1-3, pp.~100--118, 1985.

\bibitem{banerjee2020recurrence}
A.~Banerjee, B.~Goswami, Y.~Hirata, D.~Eroglu, B.~Merz, J.~Kurths, and
  N.~Marwan, ``Recurrence analysis of extreme event like data,'' {\em Nonlinear
  Processes in Geophysics Discussions}, pp.~1--25, 2020.

\bibitem{ozken2018recurrence}
I.~Ozken, D.~Eroglu, S.~F. Breitenbach, N.~Marwan, L.~Tan, U.~Tirnakli, and
  J.~Kurths, ``Recurrence plot analysis of irregularly sampled data,'' {\em
  Physical Review E}, vol.~98, no.~5, p.~052215, 2018.

\bibitem{marwan2007recurrence}
N.~Marwan, M.~C. Romano, M.~Thiel, and J.~Kurths, ``Recurrence plots for the
  analysis of complex systems,'' {\em Physics reports}, vol.~438, no.~5-6,
  pp.~237--329, 2007.

\bibitem{garcia2018classification}
E.~Garcia-Ceja, M.~Z. Uddin, and J.~Torresen, ``Classification of recurrence
  plots’ distance matrices with a convolutional neural network for activity
  recognition,'' {\em Procedia computer science}, vol.~130, pp.~157--163, 2018.

\bibitem{romano2004multivariate}
M.~C. Romano, M.~Thiel, J.~Kurths, and W.~von Bloh, ``Multivariate recurrence
  plots,'' {\em Physics letters A}, vol.~330, no.~3-4, pp.~214--223, 2004.

\bibitem{marwan2013recurrence}
N.~Marwan, S.~Schinkel, and J.~Kurths, ``Recurrence plots 25 years
  later—gaining confidence in dynamical transitions,'' {\em EPL (Europhysics
  Letters)}, vol.~101, no.~2, p.~20007, 2013.

\bibitem{corso2018quantifying}
G.~Corso, T.~d.~L. Prado, G.~Z. d.~S. Lima, J.~Kurths, and S.~R. Lopes,
  ``Quantifying entropy using recurrence matrix microstates,'' {\em Chaos: An
  Interdisciplinary Journal of Nonlinear Science}, vol.~28, no.~8, p.~083108,
  2018.

\bibitem{schinkel2007order}
S.~Schinkel, N.~Marwan, and J.~Kurths, ``Order patterns recurrence plots in the
  analysis of erp data,'' {\em Cognitive neurodynamics}, vol.~1, no.~4,
  pp.~317--325, 2007.

\bibitem{braun2021detection}
T.~Braun, V.~R. Unni, R.~Sujith, J.~Kurths, and N.~Marwan, ``Detection of
  dynamical regime transitions with lacunarity as a multiscale recurrence
  quantification measure,'' {\em Nonlinear Dynamics}, pp.~1--19, 2021.

\bibitem{marwan2018regime}
N.~Marwan, D.~Eroglu, I.~Ozken, T.~Stemler, K.-H. Wyrwoll, and J.~Kurths,
  ``Regime change detection in irregularly sampled time series,'' in {\em
  Advances in Nonlinear Geosciences}, pp.~357--368, Springer, 2018.

\bibitem{stegner2019inferring}
M.~A. Stegner, Z.~Ratajczak, S.~R. Carpenter, and J.~W. Williams, ``Inferring
  critical transitions in paleoecological time series with irregular sampling
  and variable time-averaging,'' {\em Quaternary Science Reviews}, vol.~207,
  pp.~49--63, 2019.

\bibitem{kraemer2021unified}
K.~H. Kraemer, G.~Datseris, J.~Kurths, I.~Z. Kiss, J.~L. Ocampo-Espindola, and
  N.~Marwan, ``A unified and automated approach to attractor reconstruction,''
  {\em New Journal of Physics}, 2021.

\bibitem{rehfeld2014similarity}
K.~Rehfeld and J.~Kurths, ``Similarity estimators for irregular and
  age-uncertain time series,'' {\em Climate of the Past}, vol.~10, no.~1,
  pp.~107--122, 2014.

\bibitem{mudelsee2013climate}
M.~Mudelsee, {\em Climate time series analysis}.
\newblock Springer, 2013.

\bibitem{masek1980faster}
W.~J. Masek and M.~S. Paterson, ``A faster algorithm computing string edit
  distances,'' {\em Journal of Computer and System sciences}, vol.~20, no.~1,
  pp.~18--31, 1980.

\bibitem{rabiner1978considerations}
L.~Rabiner, A.~Rosenberg, and S.~Levinson, ``Considerations in dynamic time
  warping algorithms for discrete word recognition,'' {\em IEEE Transactions on
  Acoustics, Speech, and Signal Processing}, vol.~26, no.~6, pp.~575--582,
  1978.

\bibitem{victor1997metric}
J.~D. Victor and K.~P. Purpura, ``Metric-space analysis of spike trains:
  theory, algorithms and application,'' {\em Network: computation in neural
  systems}, vol.~8, no.~2, pp.~127--164, 1997.

\bibitem{jp1987recurrence}
E.~Jp, ``Recurrence plots of dynamical systems,'' {\em Europhysics Ltters},
  vol.~5, pp.~973--977, 1987.

\bibitem{poincare1890probleme}
H.~Poincar{\'e}, ``Sur le probl{\`e}me des trois corps et les {\'e}quations de
  la dynamique,'' {\em Acta mathematica}, vol.~13, no.~1, pp.~A3--A270, 1890.

\bibitem{kraemer2018recurrence}
K.~H. Kraemer, R.~V. Donner, J.~Heitzig, and N.~Marwan, ``Recurrence threshold
  selection for obtaining robust recurrence characteristics in different
  embedding dimensions,'' {\em Chaos: An Interdisciplinary Journal of Nonlinear
  Science}, vol.~28, no.~8, p.~085720, 2018.

\bibitem{schinkel2008}
S.~Schinkel, O.~Dimigen, and N.~Marwan, ``Selection of recurrence threshold for
  signal detection,'' {\em European Physical Journal -- Special Topics},
  vol.~164, no.~1, pp.~45--53, 2008.

\bibitem{hirata2021recurrence}
Y.~Hirata, ``Recurrence plots for characterizing random dynamical systems,''
  {\em Communications in Nonlinear Science and Numerical Simulation}, vol.~94,
  p.~105552, 2021.

\bibitem{ramos2017recurrence}
A.~M. Ramos, A.~Builes-Jaramillo, G.~Poveda, B.~Goswami, E.~E. Macau,
  J.~Kurths, and N.~Marwan, ``Recurrence measure of conditional dependence and
  applications,'' {\em Physical Review E}, vol.~95, no.~5, p.~052206, 2017.

\bibitem{westerhold2020astronomically}
T.~Westerhold, N.~Marwan, A.~J. Drury, D.~Liebrand, C.~Agnini, E.~Anagnostou,
  J.~S. Barnet, S.~M. Bohaty, D.~De~Vleeschouwer, F.~Florindo, {\em et~al.},
  ``An astronomically dated record of earth’s climate and its predictability
  over the last 66 million years,'' {\em Science}, vol.~369, no.~6509,
  pp.~1383--1387, 2020.

\bibitem{kantz1994robust}
H.~Kantz, ``A robust method to estimate the maximal lyapunov exponent of a time
  series,'' {\em Physics letters A}, vol.~185, no.~1, pp.~77--87, 1994.

\bibitem{donner2008nonlinear}
R.~V. Donner and S.~M. Barbosa, ``Nonlinear time series analysis in the
  geosciences,'' {\em Lecture Notes in Earth Sciences}, vol.~112, 2008.

\bibitem{march2005recurrence}
T.~March, S.~Chapman, and R.~Dendy, ``Recurrence plot statistics and the effect
  of embedding,'' {\em Physica D: Nonlinear Phenomena}, vol.~200, no.~1-2,
  pp.~171--184, 2005.

\bibitem{williams2011finite}
D.~C. Williams, ``Finite sample correction factors for several simple robust
  estimators of normal standard deviation,'' {\em Journal of Statistical
  Computation and Simulation}, vol.~81, no.~11, pp.~1697--1702, 2011.

\bibitem{park2019investigation}
C.~Park, H.~Kim, and M.~Wang, ``Investigation of finite-sample properties of
  robust location and scale estimators,'' {\em Communications in
  Statistics-Simulation and Computation}, pp.~1--27, 2019.

\bibitem{trenberth1984some}
K.~E. Trenberth, ``Some effects of finite sample size and persistence on
  meteorological statistics. part i: Autocorrelations,'' {\em Monthly Weather
  Review}, vol.~112, no.~12, pp.~2359--2368, 1984.

\bibitem{goswami2017inferring}
B.~Goswami, P.~Schultz, B.~Heinze, N.~Marwan, B.~Bodirsky, H.~Lotze-Campen, and
  J.~Kurths, ``Inferring interdependencies from short time series,'' in {\em
  Indian Academy of Sciences Conference Series}, vol.~1, pp.~51--60, 2017.

\bibitem{schreiber2000surrogate}
T.~Schreiber and A.~Schmitz, ``Surrogate time series,'' {\em Physica D:
  Nonlinear Phenomena}, vol.~142, no.~3-4, pp.~346--382, 2000.

\bibitem{abramoff2004image}
M.~D. Abr{\`a}moff, P.~J. Magalh{\~a}es, and S.~J. Ram, ``Image processing with
  imagej,'' {\em Biophotonics international}, vol.~11, no.~7, pp.~36--42, 2004.

\bibitem{aharon2006caves}
P.~Aharon, M.~Rasbury, and V.~Murgulet, ``Caves of niue island, south pacific:
  Speleothems and water geochemistry,'' {\em Geological Society of America
  Special Papers}, vol.~404, pp.~283--295, 2006.

\bibitem{CINTHYA}
C.~Nava-Fernandez, T.~Braun, B.~Fox, A.~Hartland, O.~Kwiecien, C.~L. Pederson,
  S.~N. H{\"o}pker, S.~Bernasconi, M.~Jaggi, J.~Hellstrom, F.~Gazquez,
  A.~French, N.~Marwan, A.~Immenhauser, and S.~Breitenbach, ``Mid-holocene
  rainfall changes in the southwestern pacific,'' {\em Submitted to Climate of
  the Past}, 2021.

\bibitem{cox1967renewal}
D.~R. Cox, {\em Renewal theory}.
\newblock Springer, 1967.

\end{thebibliography}

\end{document}